\documentclass{aa}
\usepackage{graphicx}
\usepackage{txfonts}
\begin{document}

\title{Disc light variability in the FUor star V646~Puppis 
as observed by {\it TESS} and from the ground}
\titlerunning{Light variations in V646~Pup: initial study}
\authorrunning{M. Siwak, W. Og{\l}oza}
\author
{Micha{\l} Siwak\inst{1,2},
Waldemar Og{\l}oza\inst{1},
Jerzy Krzesi{\'n}ski\inst{3}
}
\institute{$^1$ Mount Suhora Observatory, Krakow Pedagogical University,
ul.\ Podchorazych 2, 30-084 Krakow, Poland\\
\email{siwak@oa.uj.edu.pl}\\
{$^2$ Konkoly Observatory, Research Centre for Astronomy and Earth Sciences, Konkoly-Thege Mikl\'os \'ut 15-17, 1121 Budapest, Hungary}\\
{$^3$ Astronomical Observatory, Jagiellonian University, ul. Orla 171, PL-30-244 Krakow, Poland}
}
\date{Received ; accepted }

\abstract
{We investigate small-scale light variations in V646~Pup occurring on timescales of days, 
weeks, and years.}
{We aim to investigate whether this variability is similar to that observed in FU~Ori.}
{
We observed V646~Pup on six occasions at the {\it SAAO} and {\it CTIO} between 2013 and 2018 with Johnson and Sloan 
filters, typically using a one-day cadence maintained for two to four weeks. 
We also utilised the public-domain 1512-day-long {\it ASAS-SN} light curve and {\it TESS} photometry 
obtained in 2019 over 24.1 days with a 30~min cadence.  
New {\it SAAO} low-resolution spectra assist in updating major disc parameters, while the archival 
high-resolution Keck spectra are used to search for temporal changes in the disc rotational profiles.
} 
{
The ground-based observations confirm the constantly decreasing brightness of V646~Pup at the rate 
of 0.018~mag yr$^{-1}$.  
Precise i-band sensitive {\it TESS} data show that the slight, $0.005-0.01$~mag, light variations imposed 
on this general trend do consist of a few independent wave trains of an apparently time-coherent nature.
Assuming that this is typical situation, based on an analysis of colour-magnitude diagrams obtained for earlier 
epochs, we were able to make a preliminarily inference that the bulk of the light changes observed could be due to the rotation of disc 
photosphere inhomogeneities, arising between 10-12~R$_{\odot}$ from the star. 
We do not exclude the possibility that these inhomogeneities could also manifest themselves in the rotational profiles of the disc, 
as obtained from the high-resolution spectra. 
Assuming Keplerian rotation of these inhomogeneities, we give a preliminary determination of the stellar 
mass at 0.7-0.9~M$_{\odot}$.
}
{
Over  certain weeks, at least, V646~Pup has shown time-coherent light variability pattern(s) that could be explained 
by the rotation of an inhomogeneous disc photosphere. 
These preliminary results are similar to those better established for FU~Ori, which suggests a common driving 
mechanism(s). 
}


\maketitle

\section{Introduction}
\label{intro}
V646~Puppis (BBW~76) is special among FU~Ori-type stars (FUors, \citealt{herbig77}) 
as it has proven to be the twin of FU~Orionis following a study by \citet{eisloffel90} and \citet{reipurth02}. 
The authors stated that the light outburst of V646~Pup apparently occurred in the 19th century, 
well before the photographic plates routinely came into
%\textbacksl\ 
use. 
Its spectrum exhibits numerous similarities with that of FU~Ori, including the intensity 
of the major absorption lines, spectral variation from F5-8~I through G0-G5-I in the visual part 
to K-M~I in the near-infrared, as well as the spectrum slope and likely the same reddening; in addition, it exhibits only small 
temporal variations in the shape of sodium and H$_{\alpha}$ lines. 
\citet{zhu08} concluded that both FU~Ori and V646~Pup have highly depleted 
or absent infalling envelopes, which implies that the bulk of the observed visual radiation does emerge directly 
from their inner discs. 

Following the indirect findings of \citet{kenyon2000}, we directly investigated the temporal 
and spectral properties of the small-scale light changes in FU~Ori using {\it MOST} satellite 
\citep{siwak13, siwak18b}. 
We demonstrated that the light curve of FU~Ori, when collected for a sufficiently long time, can
be split into particular segments, each characterised by a different variability pattern and quasi-period. 
To pinpoint mechanisms leading to this diversity, we compared the 2013-2014 ground-based and the synthetic 
colour-magnitude diagrams prepared specifically for the distinct segments. 
We found that the long-periodic family of quasi-periods could be explained 
by revolution and related changes in visibility of disc inhomogeneities localised at $\sim$15-20~R$_{\odot}$ 
from the star. 
Furthermore, we obtained that hot spots or unstable hot accretion tongues could potentially explain 
short-term (1-2~d) variability that is visible for over only one week, as the amplitude of light variations 
in $U$-filter was then observed to be twice as large as in $BVR_cI_c$ filters.\\
Encouraged by these results, we decided to initiate a similar study for V646~Pup as part of the continuation 
of our earlier work. 
To accomplish this goal, first we used a vacant time prior to observations 
of our primary targets at the South African Astronomical Observatory ({\it SAAO}) and Cerro Tololo Inter-American 
Observatory ({\it CTIO}) to check if V646~Pup shows any significant light variations occurring on the timescales 
of days and weeks. 
We also utilised precise observations obtained in 2019 by the Transiting Exoplanet Survey Satellite 
({\it TESS}). 

We describe the above and also the supplementary public-domain high-resolution Keck/HIRES spectroscopic 
and photometric All-Sky Automated Survey for Supernovae ({\it ASAS-SN}) observations in Section~\ref{observations}. 
Results of the data analysis and their immediate discussion are presented in Section~\ref{results}. 
We summarise our main findings in Section~\ref{summary}. 

% ----------------------- Fig.1 the ground and space based light curves of V646 Pup --------------
\begin{figure*}
\includegraphics[width=0.5\linewidth]{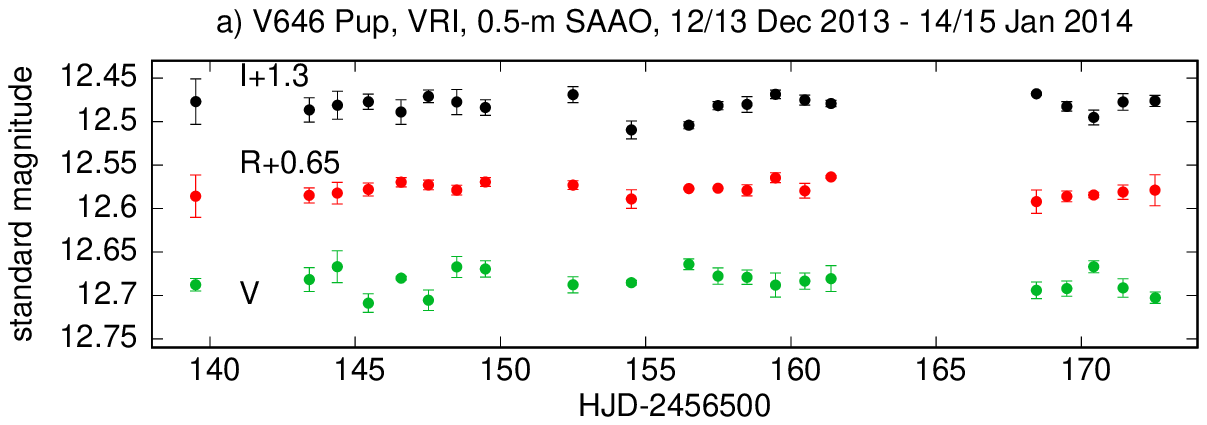}
\includegraphics[width=0.5\linewidth]{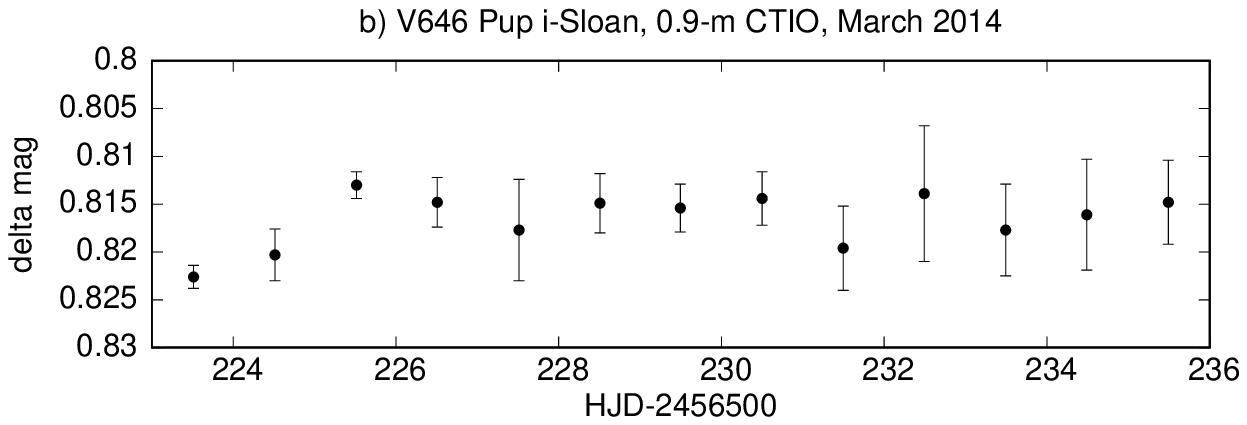}\\
\includegraphics[width=0.5\linewidth]{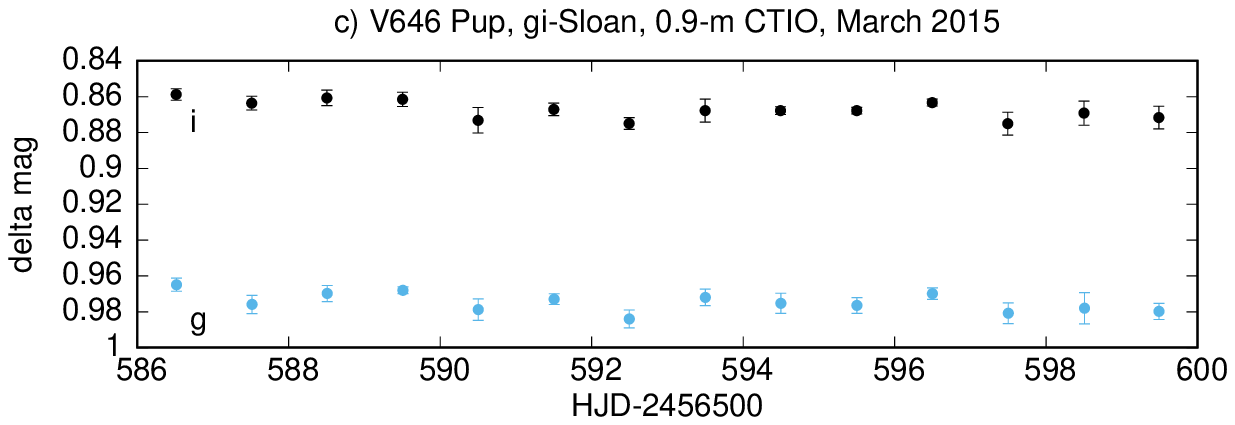} 
\includegraphics[width=0.5\linewidth]{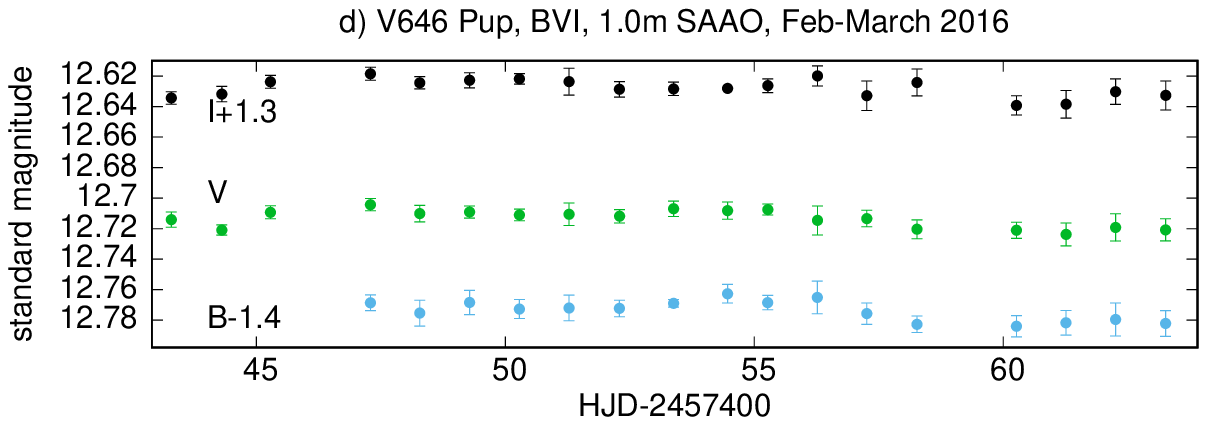}\\
\includegraphics[width=0.5\linewidth]{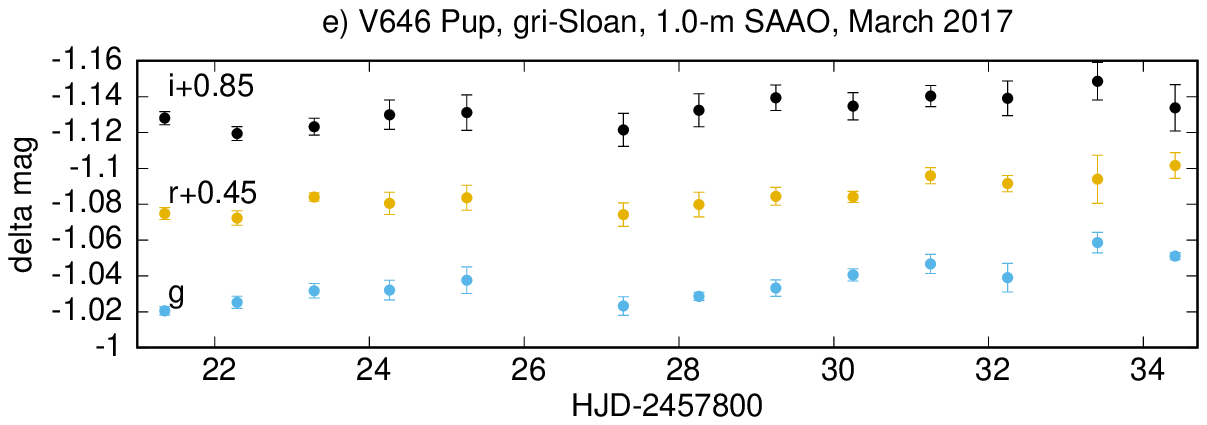}
\includegraphics[width=0.5\linewidth]{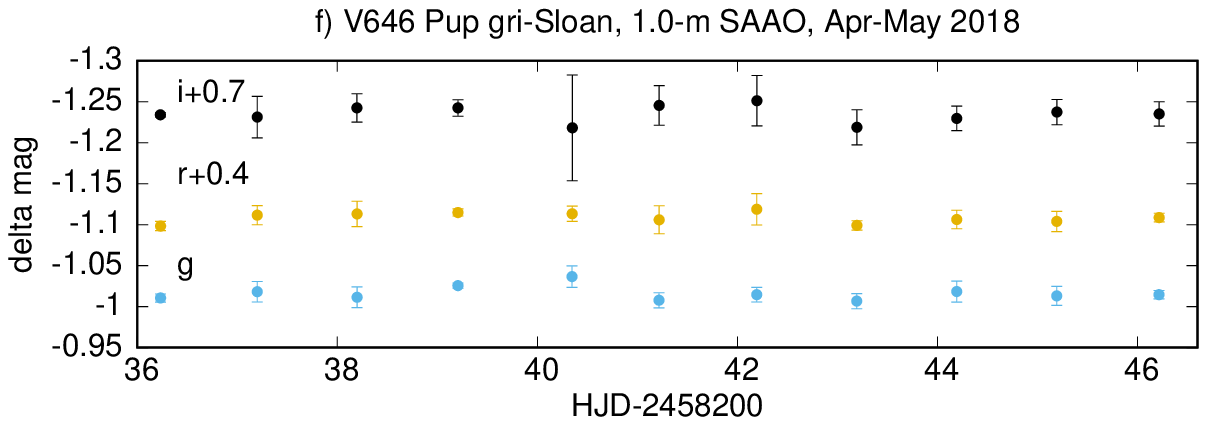}
\caption{Ground-based light curves of V646~Pup. 2013 {\it SAAO} data were standardised 
to the Johnson-Cousins system. 2016 {\it SAAO} data were approximately aligned to the standard 
Johnson system using TYC~7123-167-1 (Table~\ref{Tab.comp}).
}
\label{Fig.lcGB}
\end{figure*}
%------------------------------------------------------------------------------------------------- 

%--------- Table 1 - log of observations ---------------------------------------------------------
\begin{table*}
\caption{Log of photometric (top) and spectroscopic (bottom) observations utilised in this study.}
\begin{tabular}{c c c c c c}
\hline
Run number & Dates of observations  &  Observatory  &  Telescope/Instrument & Filters/Grating & Data file\\ \hline 
1 & 12 Dec, 2013 - 14 Jan, 2014 &  {\it SAAO} &  0.5-m / Modular Photometer & VR$_c$I$_c$     & Tab.A1\\ 
2 & 6-18 Mar, 2014              &  {\it CTIO} &  0.9-m / Tek2K~CCD          & i'              & Tab.A2\\
3 & 4-17 Mar, 2015              &  {\it CTIO} &  0.9-m / Tek2K~CCD          & g'i'            & Tab.A3\\
4 & 24 Feb - 15 Mar, 2016       &  {\it SAAO} &  1-m / STE4~CCD             & BVI             & Tab.A4\\
5 & 8-17 Mar, 2017              &  {\it SAAO} &  1-m / STE4~CCD             & g'r'i'          & Tab.A5\\
6 & 27 Apr - 7 May, 2018        &  {\it SAAO} &  1-m / STE4~CCD             & g'r'i'          & Tab.A6\\
7 & 8 Jan - 1 Feb, 2019         &  {\it TESS} &  0.1-m / Camera 3, Chip 1   & i-TESS          & Tab.A7\\ \hline
8 & 30 \& 31 Oct, 1998          &  Mauna Kea  &  Keck-I/HIRES               & UV              &  --   \\
9 &    2 Feb, 2000              &  Mauna Kea  &  Keck-I/HIRES               & RED             &  --   \\
10 & 10 \& 18 Dec, 2011         &  Mauna Kea  &  Keck-I/HIRES               & RED             &  --   \\ 
11 & 8, 12, 18 Mar, 2017        &  {\it SAAO} &  1.9-m / SpUpNIC            & gr. 6, 7        & Tab.A8\\ \hline
\end{tabular}
\label{Tab.log}
\end{table*}
%------------------------------------------------------------------------------------------------

%--------- Table 2 - magnitudes of V646 Pup, compariso and check stars---------------------------
\begin{table*}
\caption{Standard magnitudes and colour indicies of comparison stars, 
as measured at the {\it SAAO} on 11-12 Jan, 2014 \citep{siwak18b}. 
The last column contains run number(s) taken from the first column in Table~\ref{Tab.log}, 
to indicate the used comparison stars. Values labeled by 'h' are taken from \citet{hog00}.}
\begin{tabular}{c c c c c c}
\hline
stars              & $V$         & $B-V$      & $V-R_c$   & $R_c-I_c$ & Notes  \\ \hline
%{\bf V646~Pup}     & 12.689(9)   &  1.29s     & 0.753(15) & 0.752(12) & at 11/12 Jan, 2014 \\  
HD~64100           &  7.457(2)   &  1.26h     & 0.646(4)  & 0.570(4)  & \#1      \\ 
TYC~7123-167-1     & 11.677(6)   &  1.99h     & 1.110(7)  & 1.067(5)  & \#1, 2, 3  \\ 
TYC~7123-1145-1    & 11.43(9)h    &  0.35(16)h& --        & --        & \#2, 3, 4  \\     
USNOA2~0525-06468568 & $\sim13$   &  --       & --        & --        & \#2, 3    \\ 
USNOA2 0525-06466793 & $\sim14.1$ &           &           &           & \#4, 5, 6  \\
USNOA2~0525-06474617 & $\sim14.05$&           &           &           & \#4, 5, 6  \\ \hline
\end{tabular}
\label{Tab.comp}
\end{table*}
%------------------------------------------------------------------------------------------------

% ----------------------- Fig.1 ground- and space-based light curves of V646 Pup --------------
\begin{figure}
\includegraphics[width=1.0\linewidth]{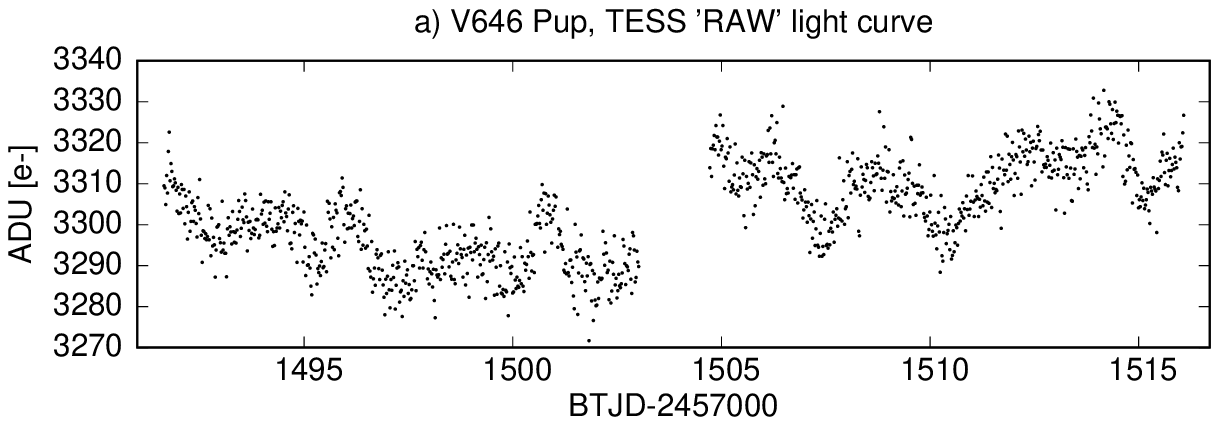}
\includegraphics[width=1.0\linewidth]{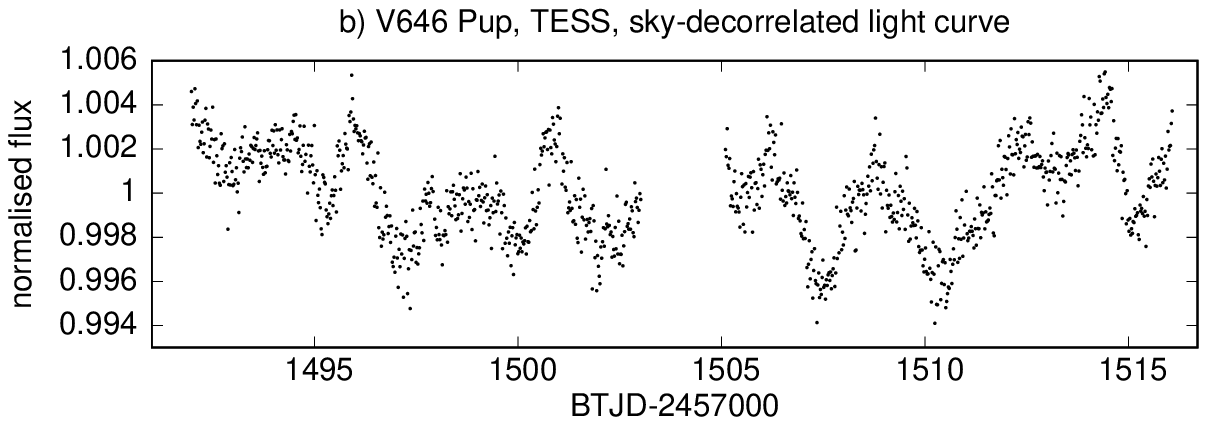}
\includegraphics[width=1.0\linewidth]{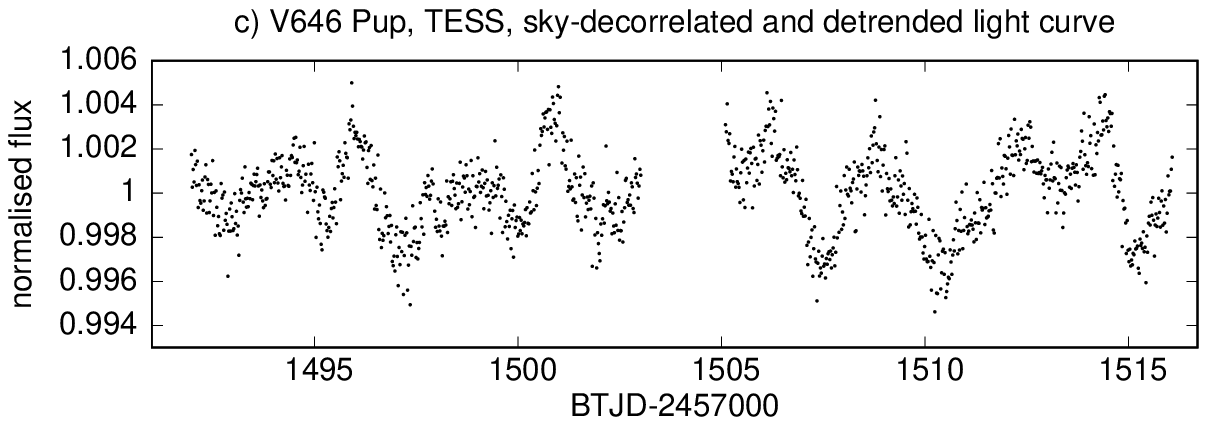}
\caption{'RAW', sky-decorrelated and detrended light curve of V646~Pup obtained by {\it TESS} in 2019.}
\label{Fig.lcTESS}
\end{figure}
%------------------------------------------------------------------------------------------------- 

\section{Observations}
\label{observations}

\subsection{Ground-based observations}
\label{multi-colour}

Photometric observations of V646~Pup with a cadence of one day were obtained on six occasions 
between December 2013 and May 2018, prior to observations of primary targets investigated 
by \citet{siwak18a, siwak18b}, whose papers contain detailed information regarding the data 
acquisition and reduction, including corrections on differential and colour extintion terms. 
Here, we only show log of these observations (Tab.~\ref{Tab.log}) and obtained light curves 
(Fig.~\ref{Fig.lcGB}, Tab.A1-A6\footnote{Tables Tab.A1-A8 are only available in electronic form
at the CDS via anonymous ftp to cdsarc.u-strasbg.fr (130.79.128.5)
or via http://cdsweb.u-strasbg.fr/cgi-bin/qcat?J/A+A/}).

\subsection{TESS observations}
\label{TESSobs}

{\it TESS} \citep{ricker15} observed V646~Pup during Cycle~1 in Sector~7. 
The full-field images (FFI) used in this work were gathered between 
3:00~UTC on 8 January to 14:00~UTC on 1 February~2019, with the cadence of 30~min. 
The total monitoring time was 24.44 days, but the run was paused after the first satellite 
orbit for 2.08~day. 
A standard aperture photometry of calibrated FFI was carried out using our {\small \sc IDL} scripts 
utilising {\small \sc DAOPHOT} \citep{stet87} procedures. 
Aperture of the 3~pix (63~arcsec) size was used for the stellar flux extraction, 
while the sky level was calculated from the annuli between 5 and 8 pixels (84-168~arcsec). 

In order to prepare the light curve for scientific analysis, first we removed 34 data points 
obtained at the beginning of each satellite orbit -- 16 in the first and 18 in the second -- which 
were heavily contaminated by scattered Earth light. 
After this operation, the total run length dropped to 24.1~d and the break in the data acquisition 
increased to 2.15~d (Fig.~\ref{Fig.lcTESS}a). 
Although the point-spread function of V646~Pup is apparently free of blending, we noticed a star-sky 
brightness correlation, which was removed by a simple linear fit (Fig.~\ref{Fig.lcTESS}b). 
Finally, we also removed the long-term trend visible in these data by fourth-order polynomial fit; the result 
is shown in Figure~\ref{Fig.lcTESS}c (Tab.A7). 
We show these results in the flux units normalised to unity at the mean star brightness level, while the time 
is Barycentric {\it TESS} Julian Day (BTJD), which is a Julian day minus 2457000, corrected to the arrival 
times at the barycentre of the Solar System.

\subsection{ASAS-SN observations}

In order to examine variability occuring on timescales of months and years, we used 1856 {\it ASAS-SN} photometric 
data points obtained in $V$- and $g'$-filters between $HJD=2457420.6652-2458385.8586$ 
and $HJD=2458283.4527-2458932.2978$, respectively. 
Details of the data acquisition, reduction, and calibration to the standard Johnson and Sloan 
systems can be found in \cite{shape14} and \cite{kochanek17}. 
Due to little brightness changes of V646~Pup, we formed 594 nightly-averages from 
%between three and four 
usually three or four 
individual data points gathered over each night. 
Their typical (median) error (standard devation) is 0.014 (0.001--0.049) and 0.016~mag (0.001-0.096~mag), 
while the typical sampling is two and one~day for the $V$- and $g'$-filter, respectively. 
The values listed in the parentheses denote the full range of errors associated with respective filters.

\subsection{Spectroscopic observations}
\label{spec-saao}

Over three nights in March, 2017, we obtained several low-resolution spectra of V646~Pup  
using the Spectrograph Upgrade Newly Improved Cassegrain ({\it SpUpNIC}), mounted 
on the 1.9-m {\it Radcliffe} telescope at the {\it SAAO} \citep{crause16}.
Gratings~7 and 6 were used to cover the wavelength ranges from 3498 to 9245~\AA ~and 
3904 to 6650~\AA ~with formal resolutions of 2.8 and 1.35~\AA~pix$^{-1}$, respectively. 
Two spectrophotometric standard stars, LTT~2415 and LTT~3864, were observed immediately 
after V646~Pup through the same slit.
The spectra were reduced on 
%calibrated with 
{\it bias} and {\it flatfield}, extracted, and then wavelength- and 
flux-calibrated ($\pm 8\%$) within the {\small \sc IRAF} package (Tab.A8).

We also requested the RAW spectra Keck archive, gathered for V646~Pup 
between 1998-2011 by HIRES \citep{vogt94}. 
Each season, the data were obtained in different spectral regions and with different spectral resolutions, 
that is, 45\,000 in 1998 and 2000, and 34\,000 in 2011. 
We routinely processed these spectra using {\it ccdproc} and {\it echelle} tasks within 
the {\small \sc IRAF} package.
 
% ----------------------- Fig.3: long term V + g light curve -------------------------------------------
\begin{figure}
\includegraphics[width=1\linewidth]{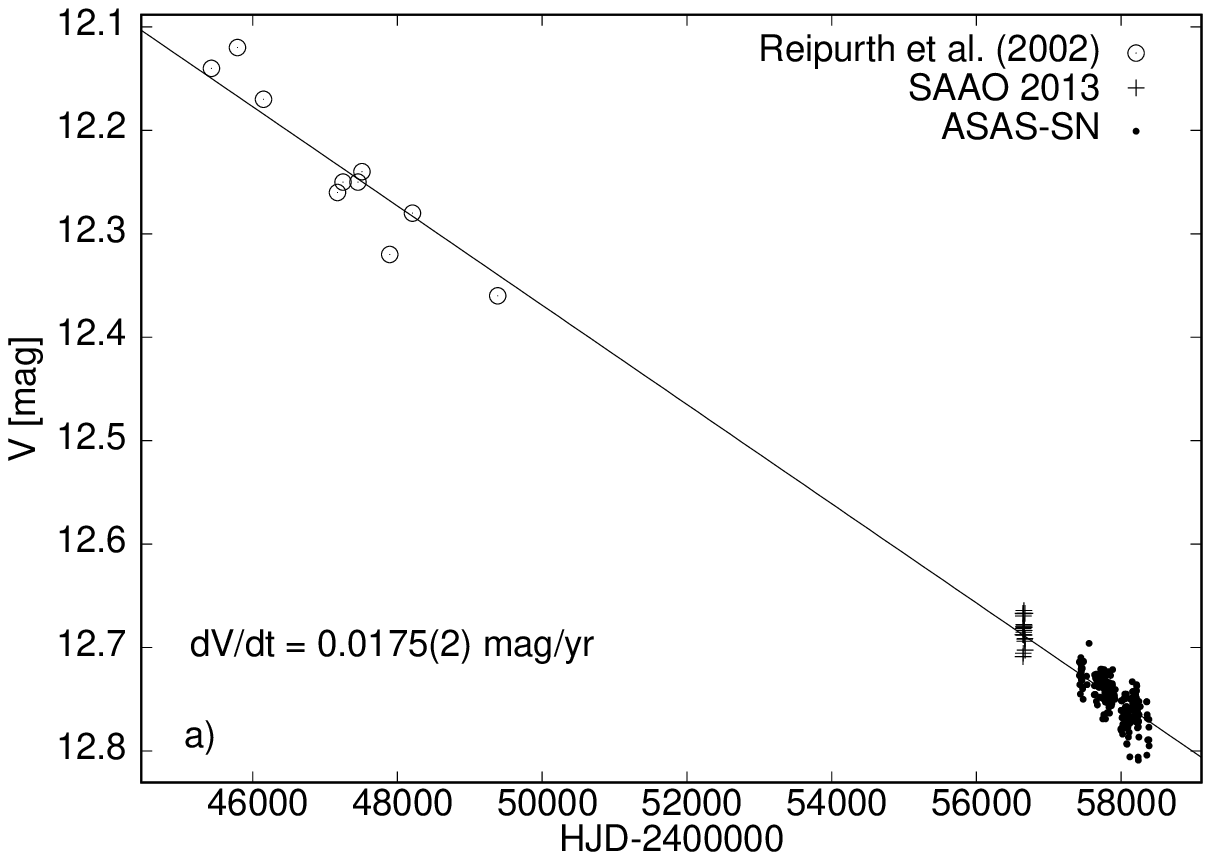}
\includegraphics[width=1\linewidth]{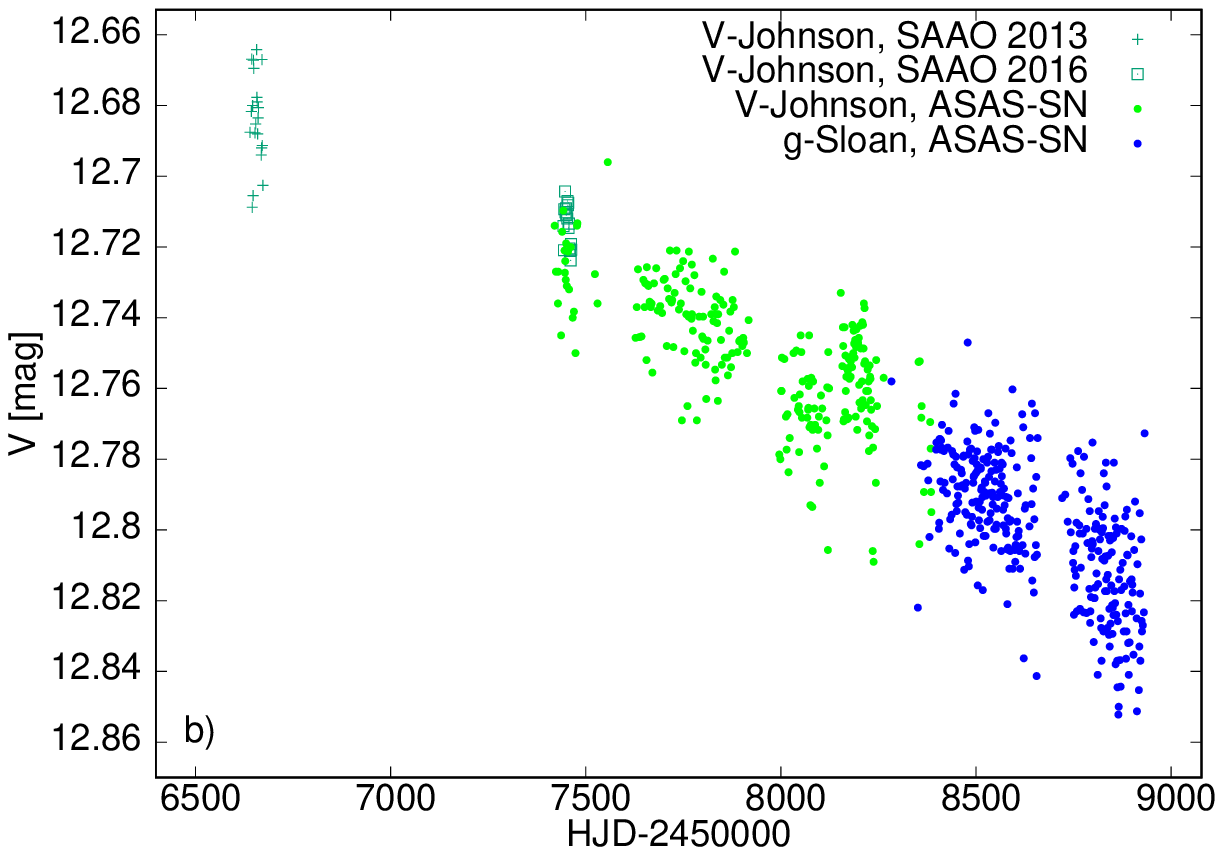}
\caption{Long-term brightness evolution of V646~Pup in $V$-filter only (panel a) and brightness 
evolution in 2013-2020 (panel b).
} 
\label{Fig.longterm}
\end{figure}
%--------------------------------------------------------------------------------------------------- 

\section{Results of the data analysis}
\label{results}

%----------------------- Fig.4: power spectrum of V -data ------------------------------------------
\begin{figure}
\includegraphics[width=1\linewidth]{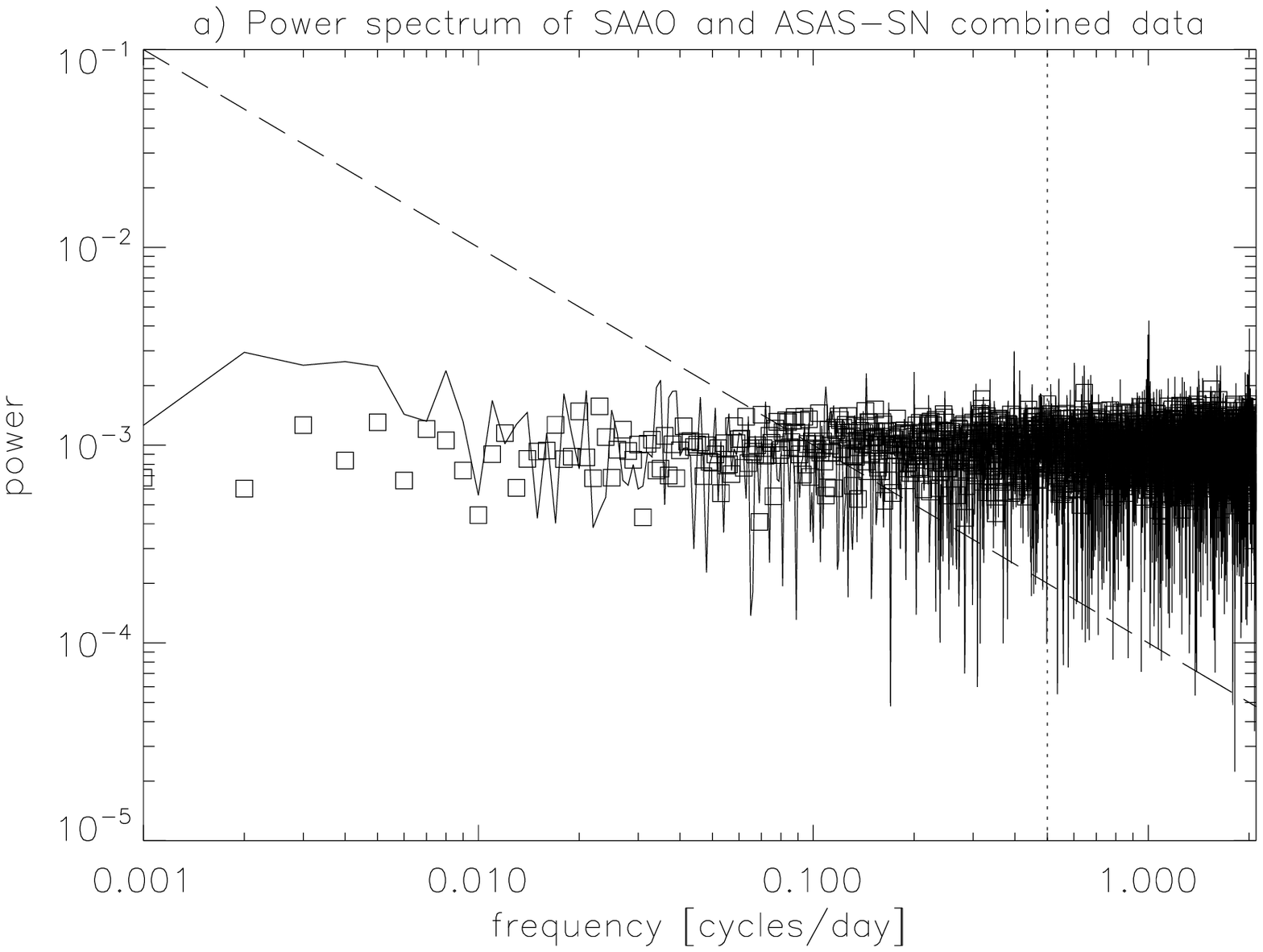}
\includegraphics[width=1\linewidth]{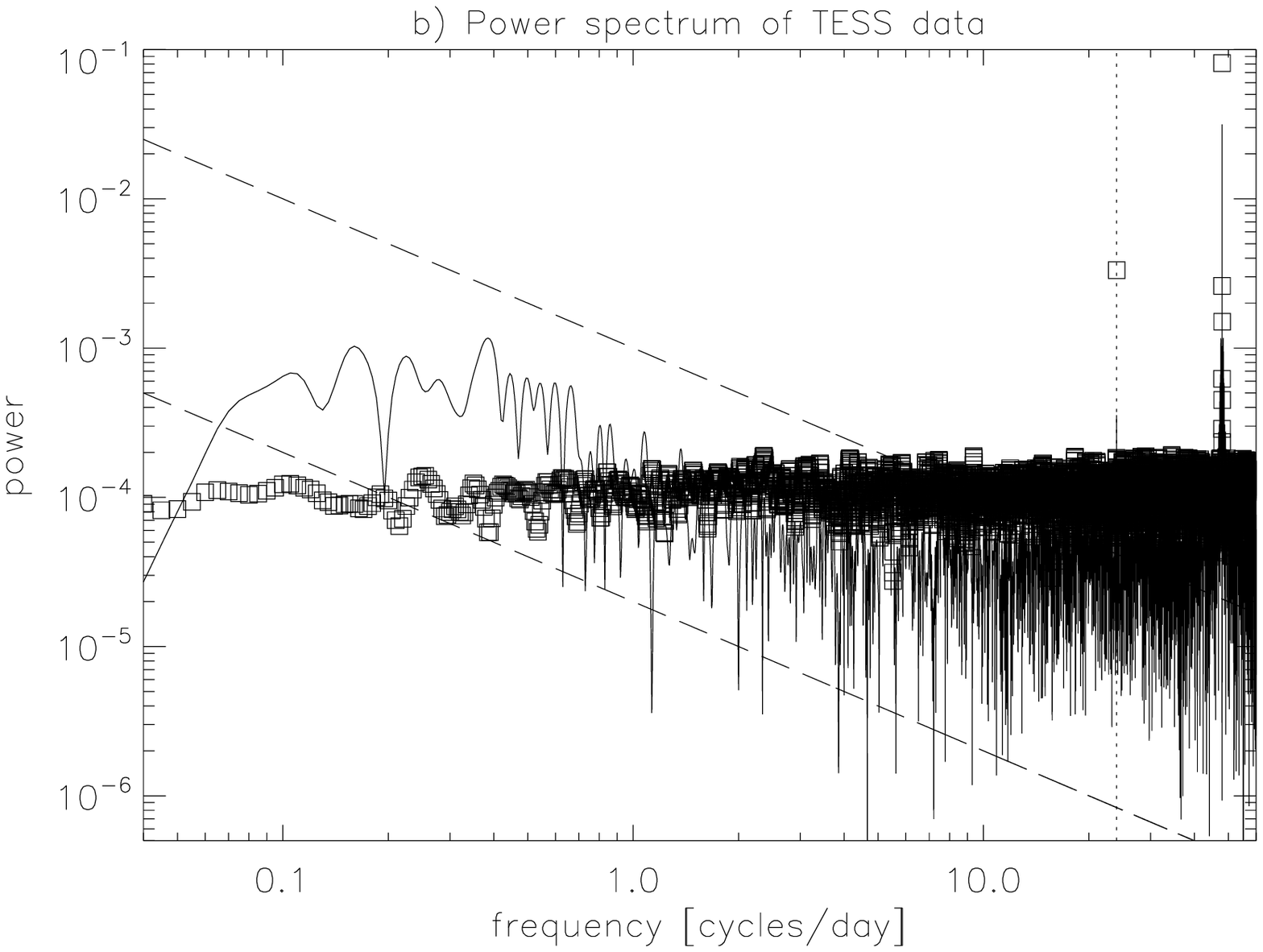}
\caption{Power spectra (solid lines) in log-log scale of combined $V$ and $g$-filter {\it SAAO} and {\it ASAS-SN} 
data (upper panel) and {\it TESS} data (bottom panel). 
The errors are marked by squares. A red flicker-noise trend, i.e. $power \sim frequency^{-1}$, is indicated by dashed 
lines for comparison. Nyquist frequencies are marked by vertical dotted lines.} 
\label{Fig.power}
\end{figure}
%--------------------------------------------------------------------------------------------------- 

\subsection{Variability seen from the ground}
\label{g-bvar}

Archival \citep{reipurth02} and the new {\it SAAO} and {\it ASAS-SN} $V$-filter data confirm 
the long-term trend of the disc light decrease (Fig.~\ref{Fig.longterm}a). 
Monitoring extended to 35 years allows us to refine the historical value of 0.023~mag~yr$^{-1}$ 
\citep{reipurth02} to $0.018$~mag~yr$^{-1}$.

We also performed a Fourier analysis to search for possible quasi-periodic oscillations (QPOs) 
in the recently gathered $V$ and $g'$-filter standardised data (the last one were combined with 
the $V$-filter data by means of Eq. 23 in \citealt{fukugita96}).  
This timeseries spans 2293 days and was obtained with one to two day cadence, except for the seasonal breaks. 
First, we removed the overall trend in magnitudes (Fig.~\ref{Fig.longterm}b) by a second-order polynomial 
fit and then transformed the obtained 'delta series' to flux units normalised to unity at the mean 
brightness level. 
We did not find any significant peaks other than those related to the breaks in the data acquisition. 
The mean standard errors of the amplitudes were calculated with bootstrap sampling technique \citep{ruc08}. 
Although the power spectrum (Fig.~\ref{Fig.power}a) may appear to show a white-noise character, this 
is a false result that is due to the small range of observed light variations, which are mostly hidden 
in the Poisson noise that dominates in the ground-based data.

In accordance with the previous finding, no unambiguous QPOs were found during the frequency analysis 
of separate and combined data sets gathered at the {\it SAAO} and {\it CTIO} between 2013-2018. 
Nonetheless, a visual inspection of individual panels in Fig.~\ref{Fig.lcGB} indicates that marginally significant 
7-8~d and 5-6~d QPOs did presumably occur in 2016 and 2017, respectively. 

% ----------------------- Fig.5: 1st part of TESS light curve, phased ------------------------------
\begin{figure*}
\includegraphics[width = 0.5\linewidth]{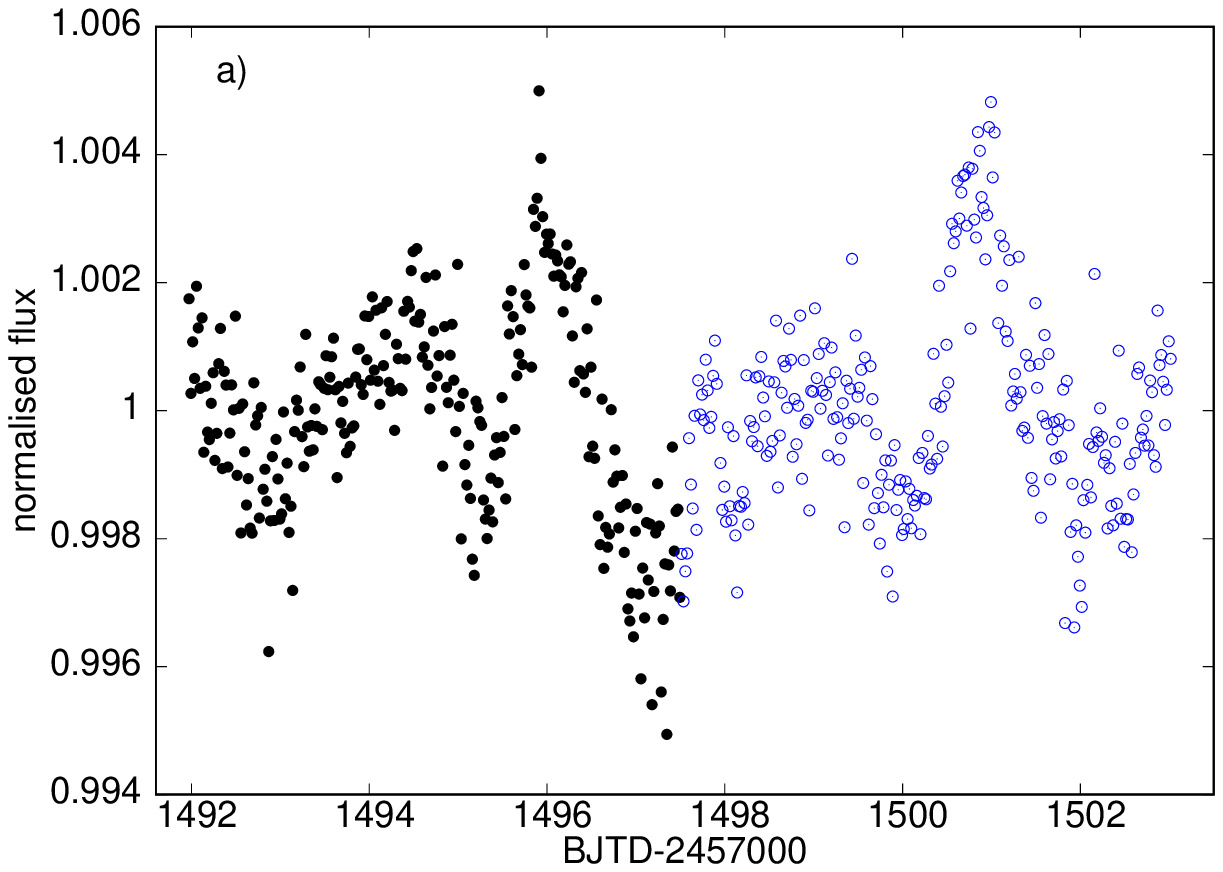}
\includegraphics[width = 0.5\linewidth]{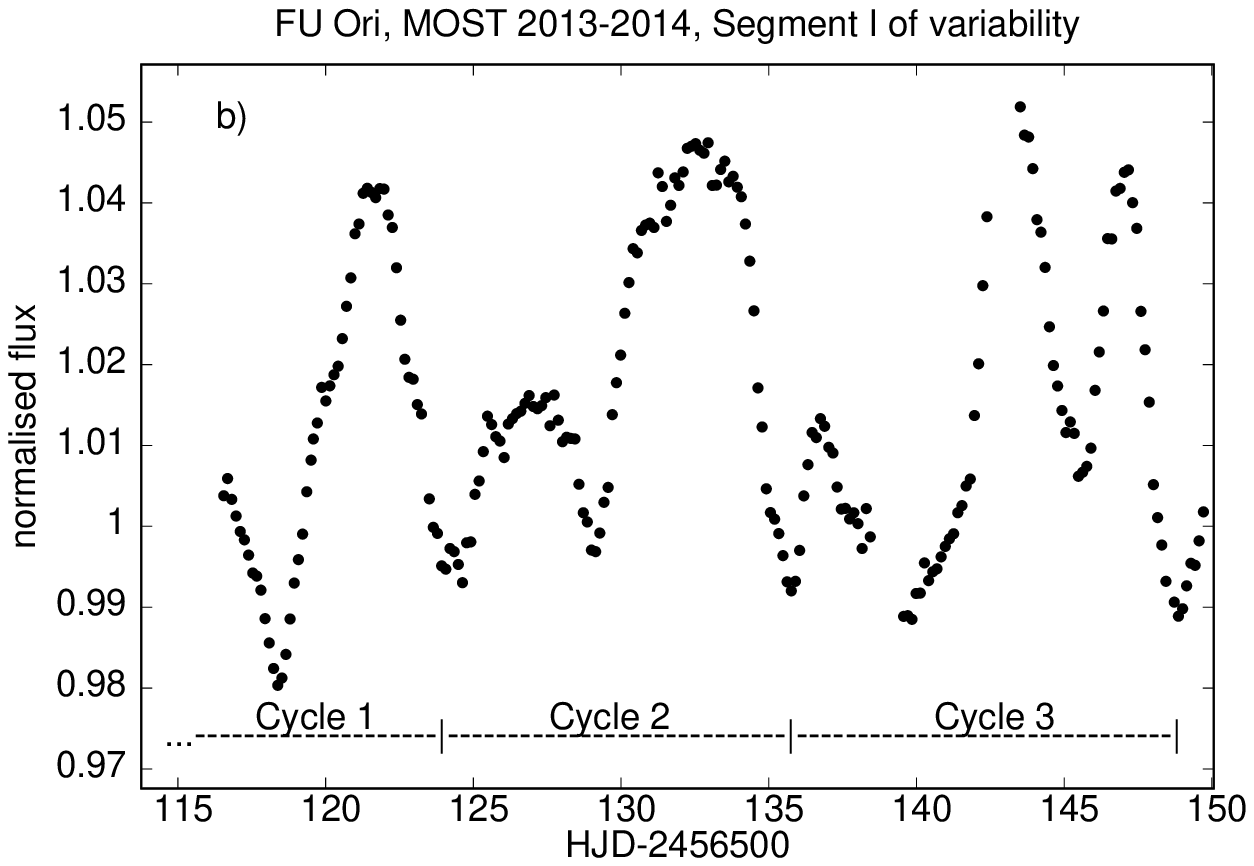}
\includegraphics[width = 0.5\linewidth]{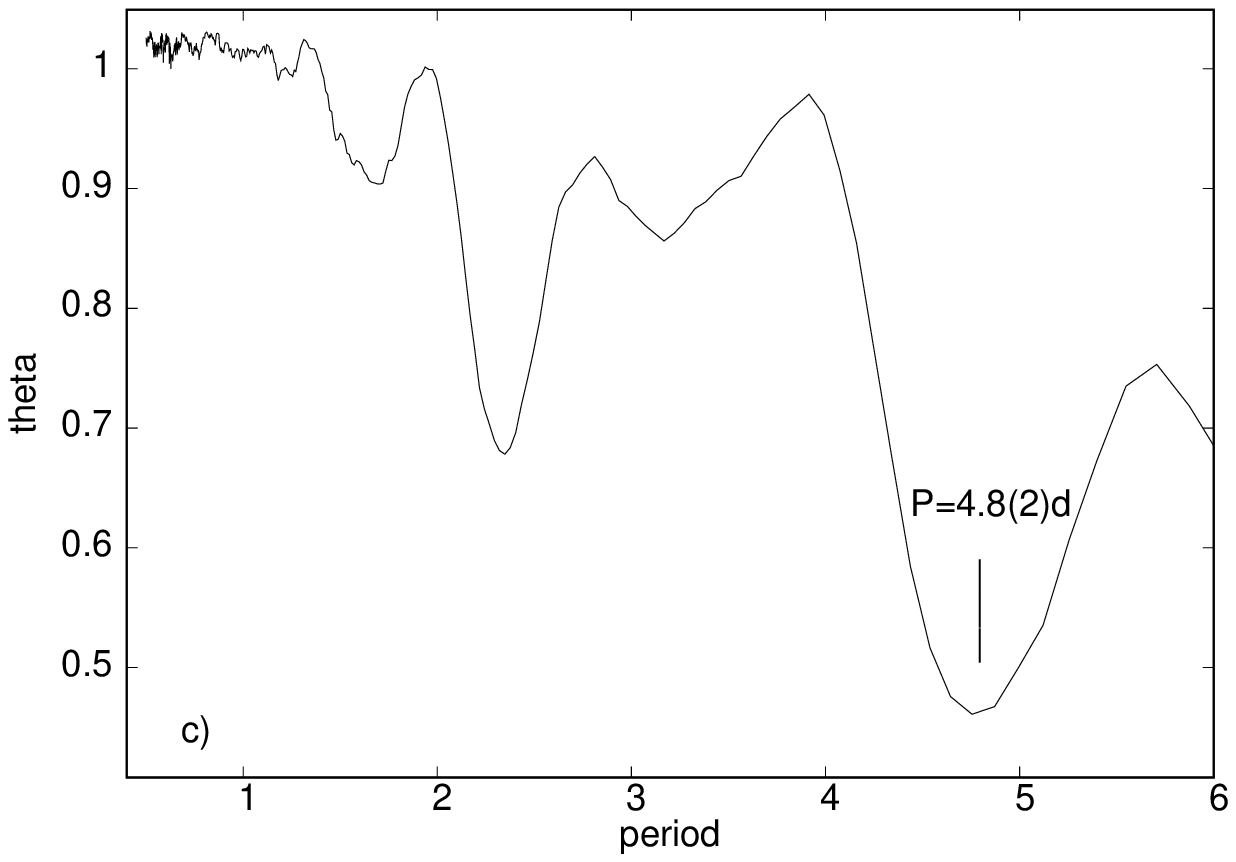} 
\includegraphics[width = 0.5\linewidth]{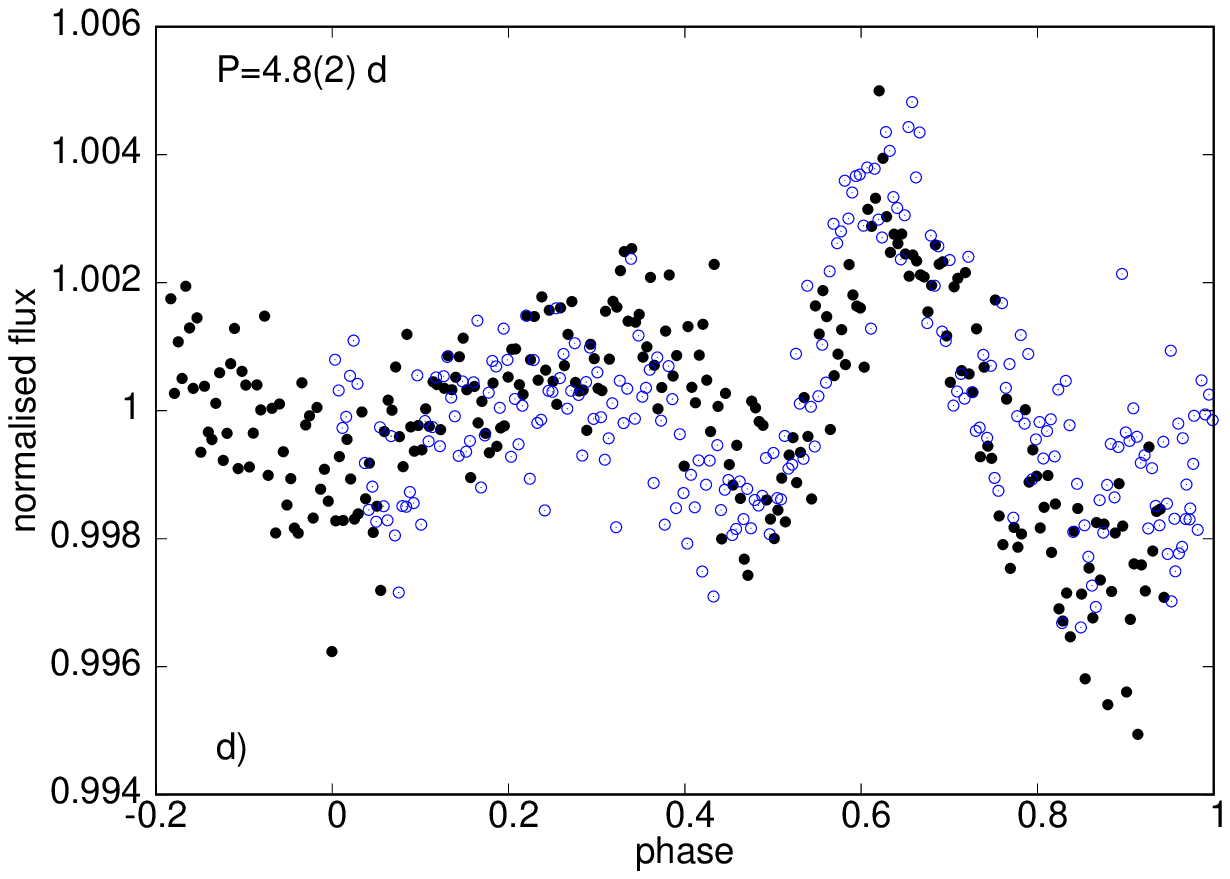}
\caption{Light curve of V646~Pup gathered during the first {\it TESS} orbit, with consecutive 
variability patterns indicated by different symbols (panel a). It is similar to FU~Ori observed by {\it MOST} 
in 2013-2014 (b). Phase dispersion minimanisation period search based on $\theta$ statistics indicates 
$4.8\pm0.2$~day period (c). Last panel (d) shows the same data as the first panel, phased with 
the above value.}
\label{Fig.phase}
\end{figure*}
%--------------------------------------------------------------------------------------------------- 

% ----------------------- Fig.6: wavelet spectrum of TESS data -------------------------------------
\begin{figure}
\includegraphics[width = 1\linewidth]{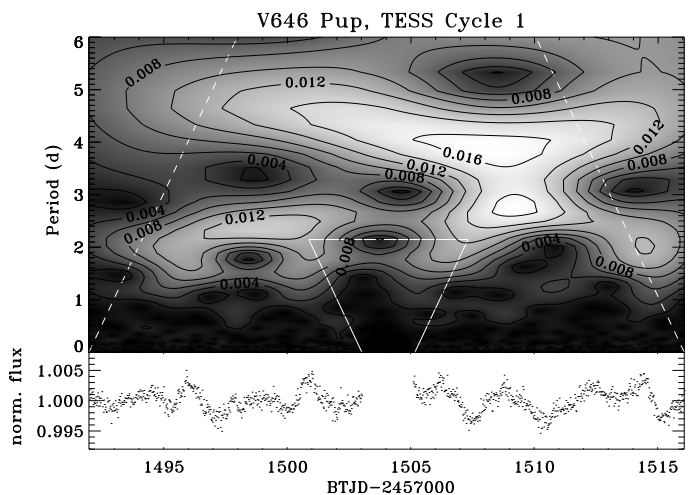} 
\caption{Wavelet spectrum of the detrended (Fig.~\ref{Fig.lcTESS}c) light curve. 
The amplitude is expressed by the grey-scale and solid black lines with values. 
The white continuous trapezium lines indicate region completely distorted by the break in the data 
acquisition (although the affected area does likely propagate to longer periods, inside the cone 
defined by the virtual extension of the trapezium). 
Edge effects are contained outside the white dashed lines.}
\label{Fig.wav}
\end{figure}
%--------------------------------------------------------------------------------------------------- 

\subsection{Variability seen by {\it TESS}}
\label{shorttermvar-tess}

Fourier analysis of {\it TESS} data obtained 
typically with the 30~min cadence (Fig.~\ref{Fig.lcTESS}c) shows a group of significant peaks 
between 0.06-0.7 cycles~d$^{-1}$, that is, $\sim1.4-5$ days (Fig.~\ref{Fig.power}b).
Thanks to limited photometric noise, the power spectrum revealed the red-noise character, as in FU~Ori. 
The {\it TESS} run length of 24.1~days stands in the way of an investigation of the low-frequency part of the spectrum. 
Nevertheless, if it is of a rotationally driven nature, these low frequencies are expected 
to arise at larger distances from the star and should best manifest themselves on infrared wavelenghts. 
This may explain why no significant long-term light variations superimposed on the general downward 
trend are seen in {\it ASAS-SN} data (Sec.~\ref{g-bvar}).

The most important information, however, is recorded in the morphology of the light curve itself:
its first part shows a repetitive pattern of light variations 
(Fig.~\ref{Fig.phase}a), which shows a striking similarity to the double-peaked pattern observed 
in FU~Ori in {\it Segment~I} by {\it MOST} (Fig.~\ref{Fig.phase}b, see also \citealt{siwak18b}).
The phase dispersion minimisation technique indicates at the $4.8\pm0.2$~d period 
evidence of this light feature in V646~Pup (Fig.~\ref{Fig.phase}c) and the phased light curve is shown 
in Fig.~\ref{Fig.phase}d. 
Its similarity to the event in FU~Ori suggests the same driving mechanism: based on the results obtained 
for FU~Ori, we induce that the major peak (at phase 0.6 for V646~Pup) could be owed to the changing visibility 
of a hot plasma bubble rotating in the disc, while the secondary wide maximum (phase 0.1-0.4) could be caused by 
the light reflected from the opposite side of the disc. 
If that is the case, assuming Keplerian rotation and the most likely stellar mass, $M$ of 0.3-0.9~M$_{\odot}$, 
the hot plasma bubble would be localised at the distance of 8-12~R$_{\odot}$, respectively. 
The lack of simultaneous multi-colour photometry does not allow for an independent confirmation of this value 
as for FU~Ori.\\
At first glance, variability during the second {\it TESS} orbit appears to be time-incoherent. 
This is also formally confirmed by means of Fourier, autocorrelation, and phase dispersion minimisation techniques. 
This ambiguity can be solved by means of the wavelet analysis utilising Morlet-6 as the mother function 
(Fig.~\ref{Fig.wav}): the wavelet 
amplitude-spectrum indicates that variability observed during the second orbit may be (at least in part) 
treated as a continuation of the double-peaked pattern. 
The wavelet spectrum also suggests period shortening similar to that observed in FU~Ori \citep{siwak13,siwak18b}. 
This phenomenon was also seen in ordinary CTTS such as TW~Hya \citep{siwak18a} 
and RU~Lup \citep{siwak16}, admittedly not directly in the disc light, but via hot spots formed 
on the stellar photospheres by unstable accretion tongues that directly reflect the temporary inner disc dynamic. 
Unfortunately, the information loss during the 2.15~d gap prevents any tracking of the evolution of the initial 
4.8~d QPO with better confidence: it probably ceased at BTJD=2458510 as a 3.8-4~d quasi-period. 
In addition, the wavelet spectrum shows a significant brightening that appeared 
at BTJD=2458507 for only 2.7-3~d and disturbed the former double-peaked pattern. 
The other possibility is that amplitudes of these two peaks were evolving over time, as inferred 
for {\it Segment~I} in FU~Ori (see Sec.~3.1 in \citealt{siwak18b}). 

% ----------------------- Fig.7: colour-magnitude diagrams ------------------------------
\begin{figure*}
\includegraphics[width=0.33\linewidth]{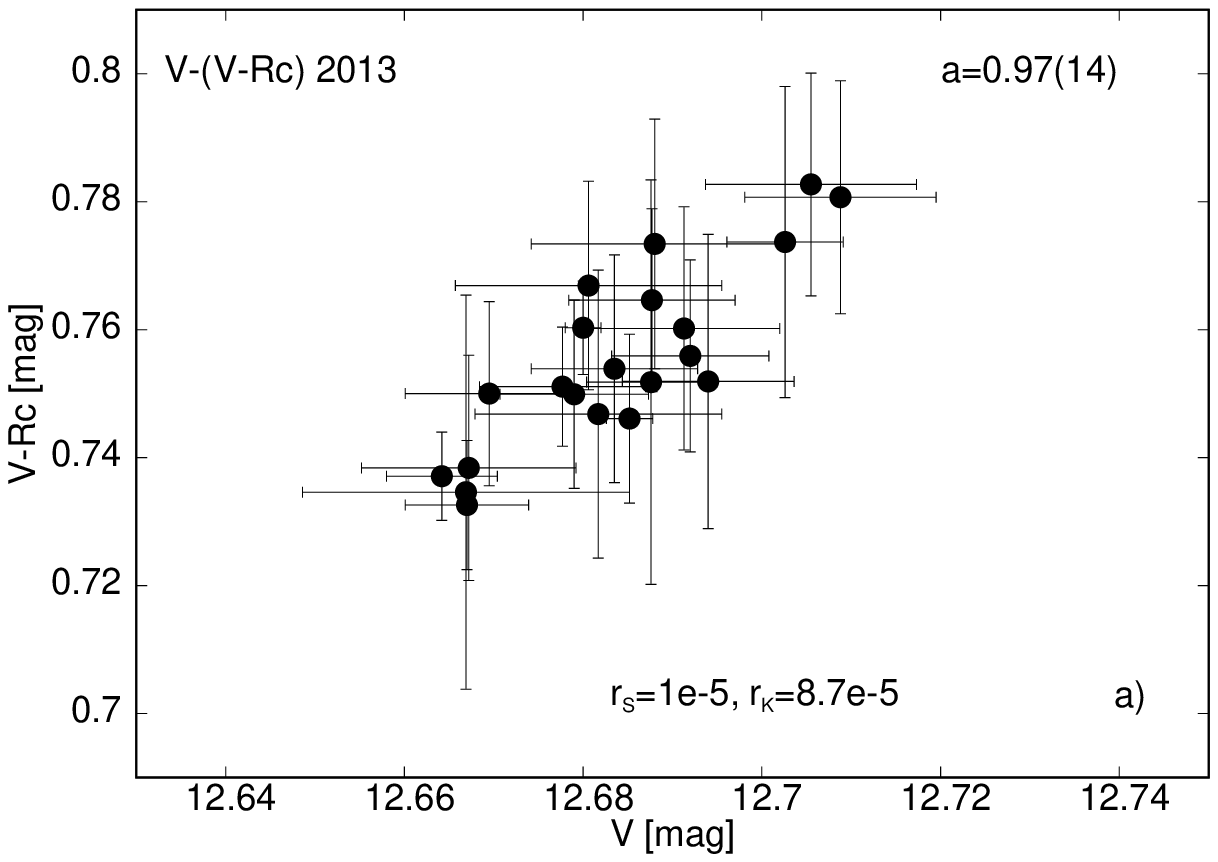}
\includegraphics[width=0.33\linewidth]{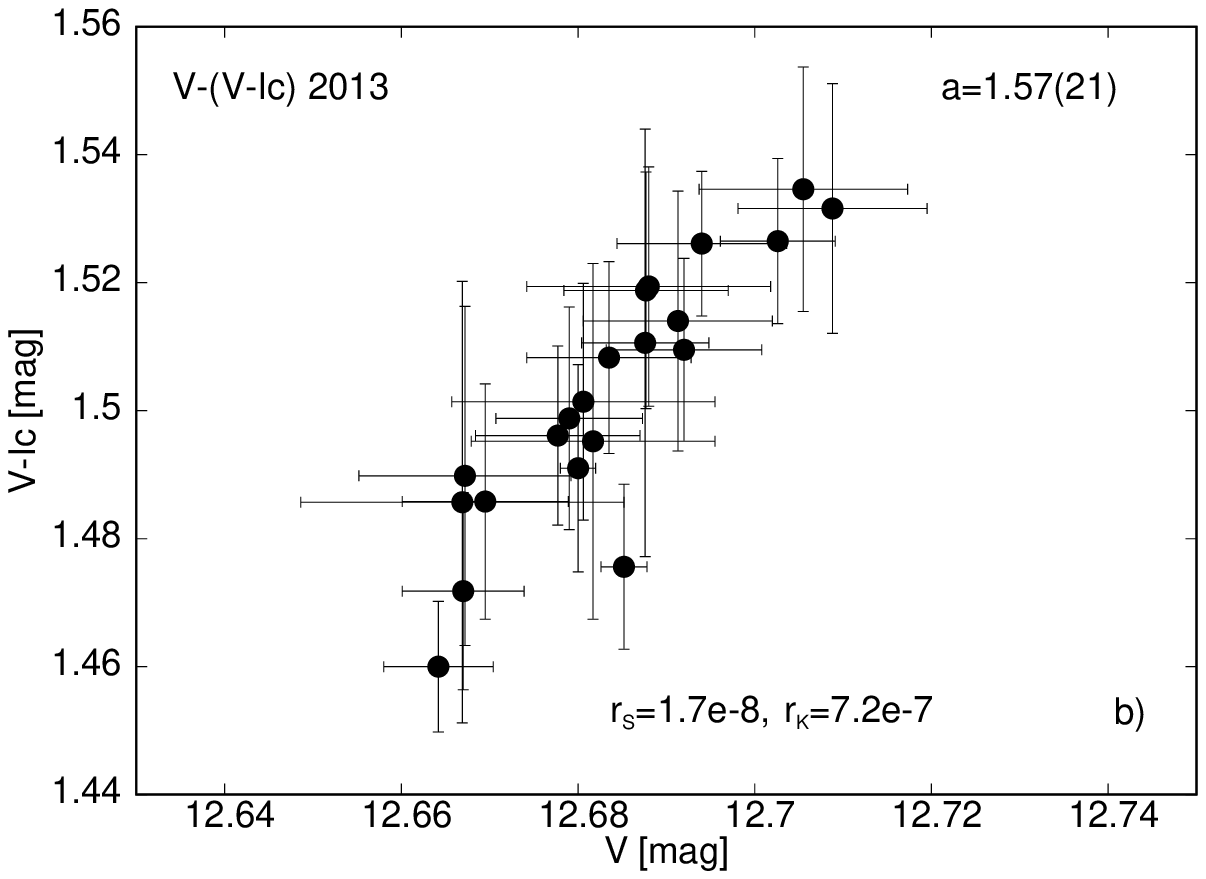}
\includegraphics[width=0.33\linewidth]{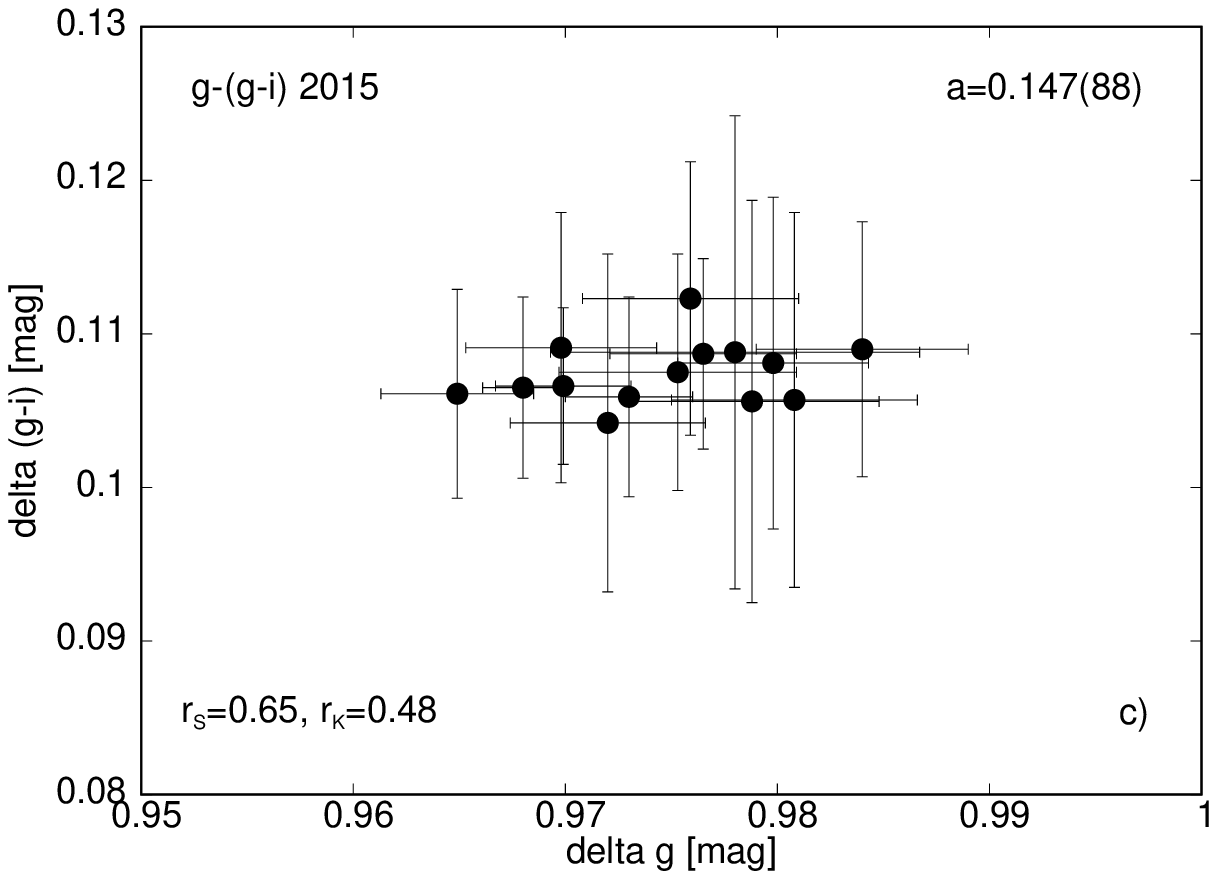}
\includegraphics[width=0.33\linewidth]{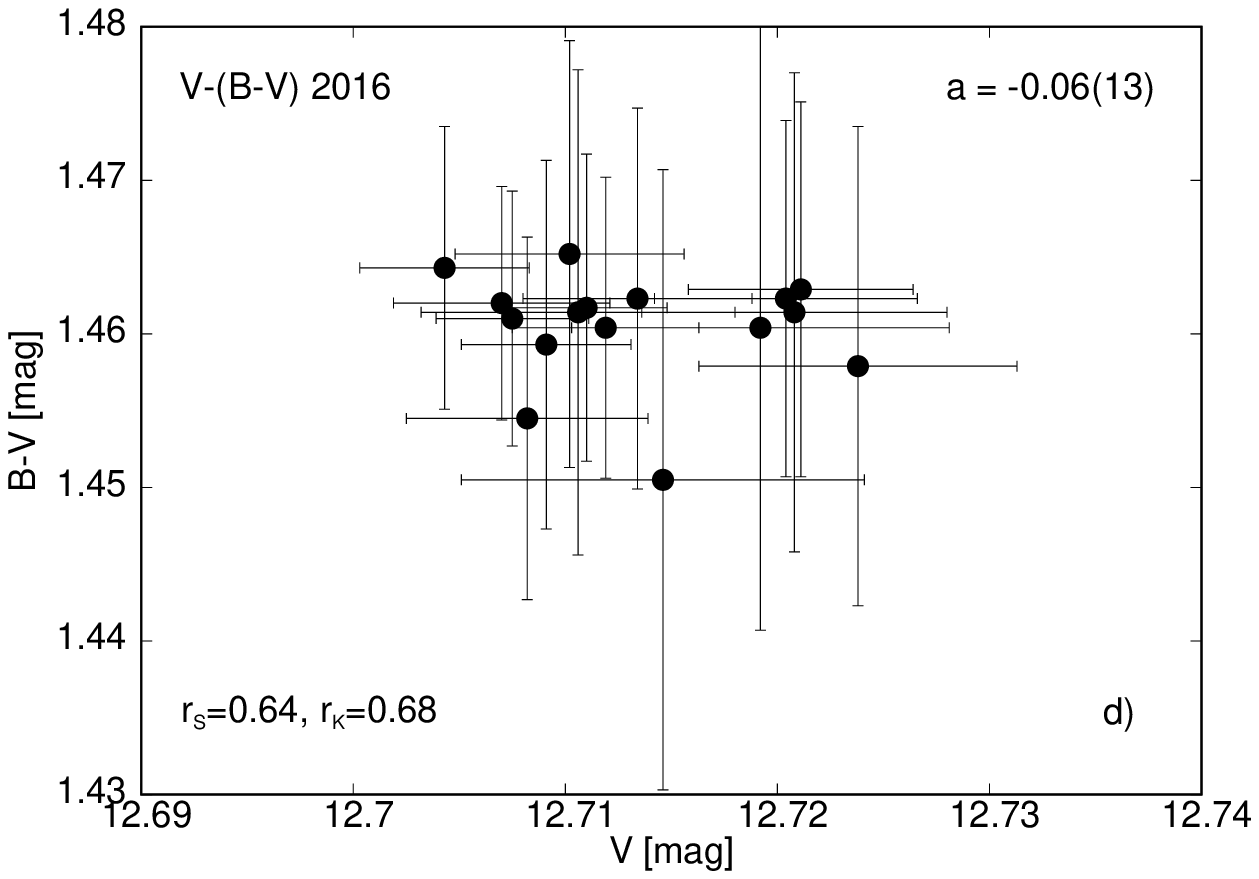}
\includegraphics[width=0.33\linewidth]{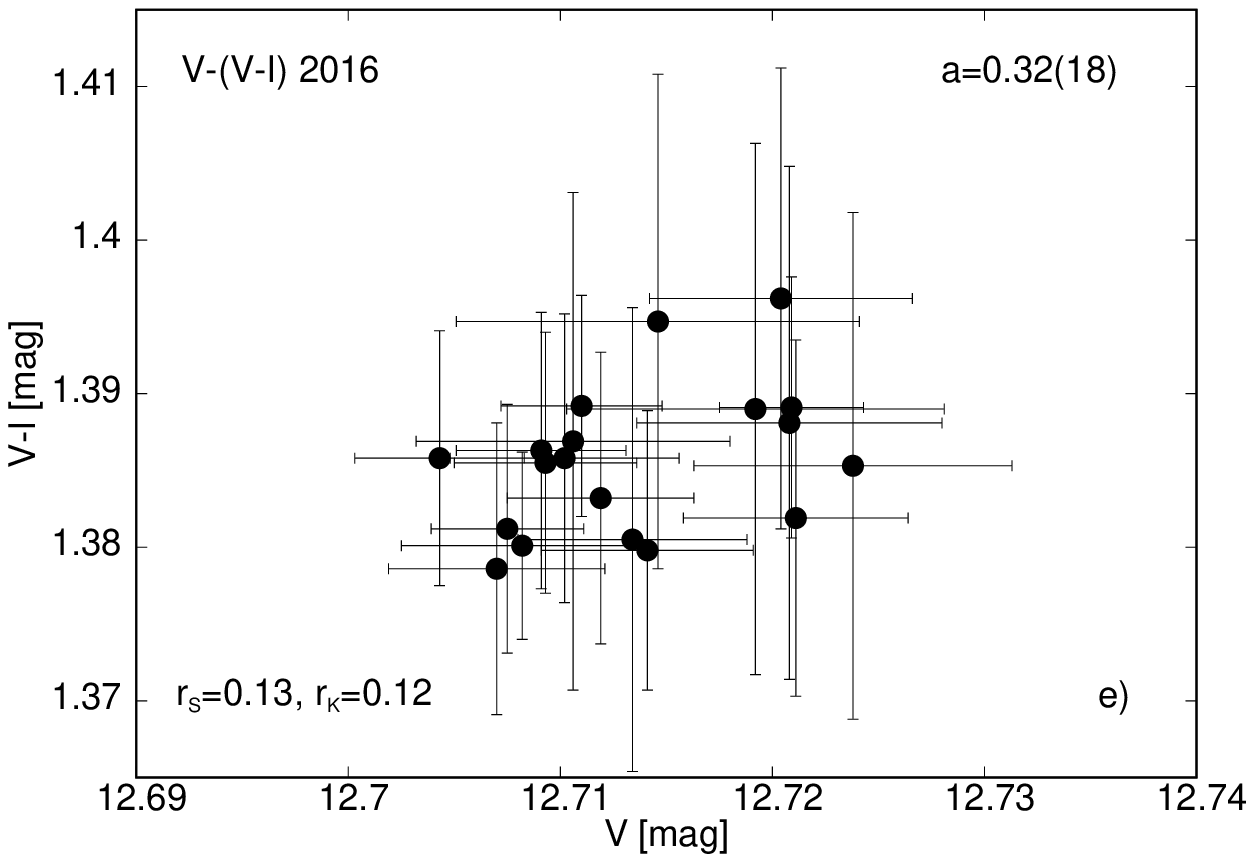}
\includegraphics[width=0.33\linewidth]{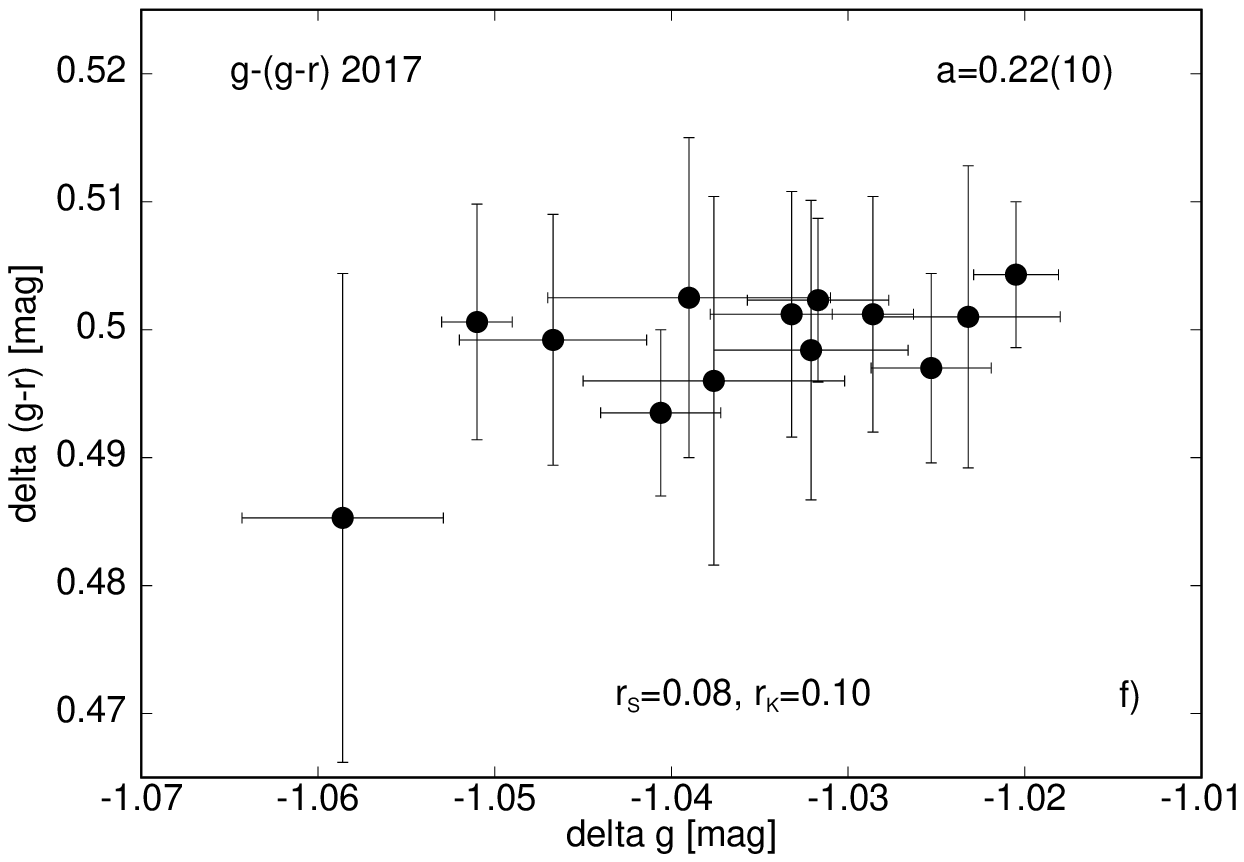}
\includegraphics[width=0.33\linewidth]{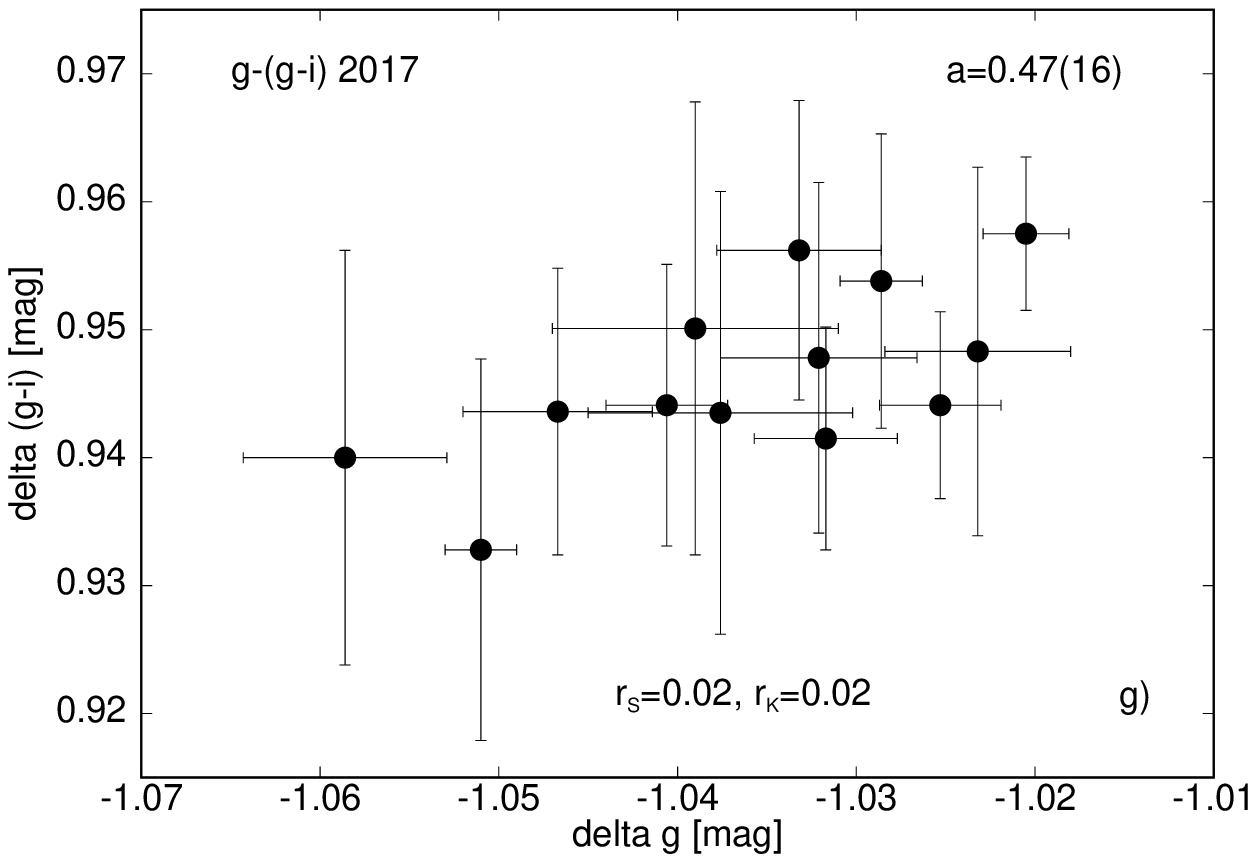}
\includegraphics[width=0.33\linewidth]{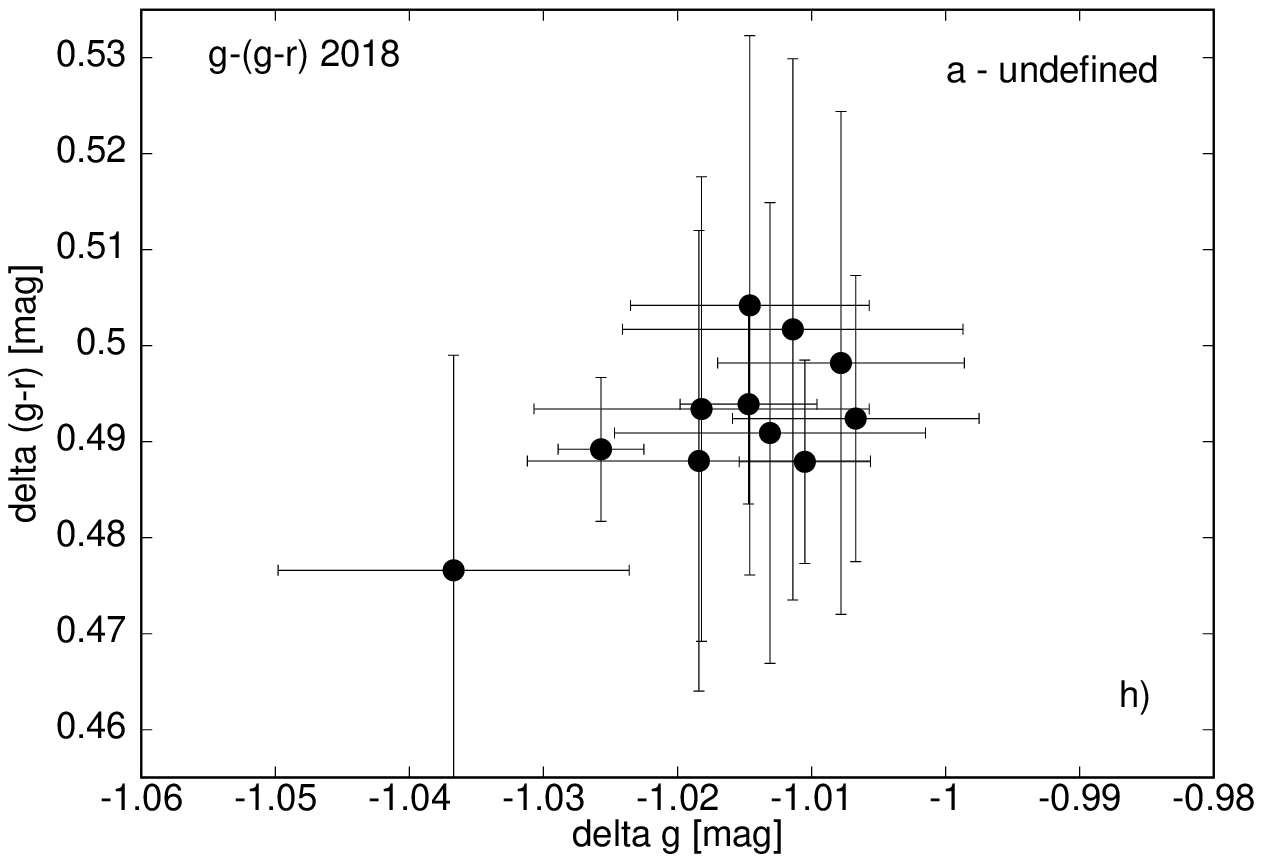}
\includegraphics[width=0.33\linewidth]{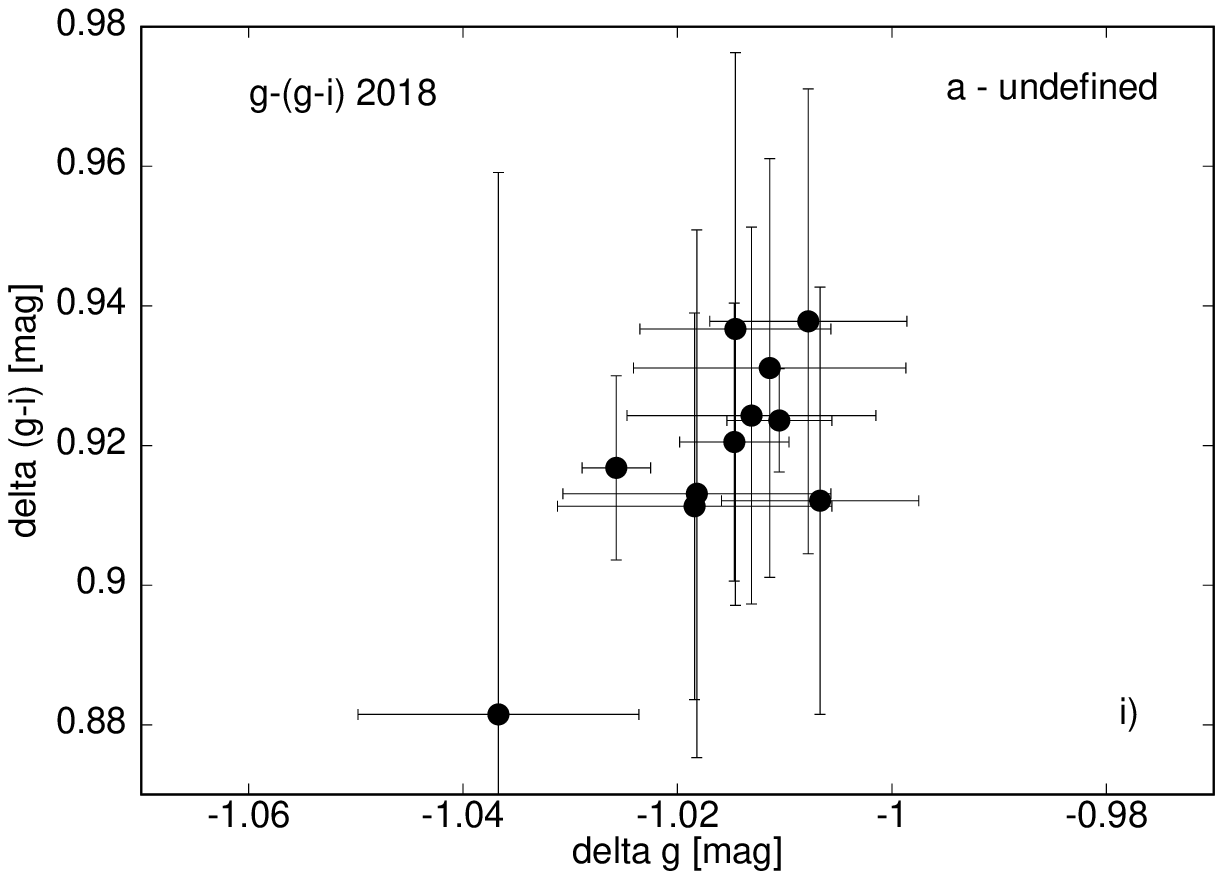}
\caption{Colour-magnitude diagrams; The only significant trends are seen in 2013, 2016 and 2017. 
Their errors are in parentheses.} 
\label{Fig.shortterm}
\end{figure*}
%----------------------------------------------------------------------------------------

\subsection{Colour-index variations}
\label{CIvar}

The variety of photometric systems used in this work and frequent lack of standarisation 
prevents studies of colour index variations occurring on long timescales. 
For this reason this investigation is limited to variations occurring during separate runs only.

The most obvious correlations are seen in the colour-magnitude diagrams constructed 
from 2013 {\it SAAO} data (Fig.~\ref{Fig.shortterm}a,b). 
We give the value of the slope $a$ obtained by least squares fits weighted 
by photometric errors, as well as Sperman's $r_S$ and Kendall's $r_K$ correlation 
ranks to indicate the goodness of these fits. 
Interestingly, no QPO is seen in the associated light curves. 
The 2015 diagram (Fig.~\ref{Fig.shortterm}c) shows a possible correlation, but it is of a small significance.  
No associated QPO was inferred from the respective light curves either.\\
Although the 2016 $V- (B-V)$ diagram shows a reversed colour index (Fig.~\ref{Fig.shortterm}d), 
the formal significance of this negative trend is trivial. 
However, the associated $V- (V-I)$ diagram shows more significant and -- interestingly -- the positive 
trend (Fig.~\ref{Fig.shortterm}e).
This is reminiscent of the situation observed in colour-magnitude diagrams of FU~Ori, prepared 
for a long-periodic (10-11~d) family of QPOs \citep{siwak18b}. 
Interestingly, only a slightly shorter, possibly 7-8~d QPO, was inferred from the respective light curves 
for V646~Pup (Sec.~\ref{g-bvar}).\\
The correlations in 2017 data are also significant and are associated with a possible 5-6~d QPO (Sec.~\ref{g-bvar}). 
In contrast, no significant correlations nor QPOs are observed in 2018 (Fig.~\ref{Fig.shortterm}h,i).

\subsection{Light synthesis model versus ground-based photometry}
\label{discmodel}

Encouraged by the purely phenomenological findings of light variability 
pattern in V646~Pup that is similar to that observed in FU~Ori (Sec.~\ref{shorttermvar-tess}), 
and signs of similar trends in colour-magnitude diagrams (Sec.~\ref{CIvar}), we induce that 
variability of these two discs may be driven by common physical mechanisms. 
These variations can be preliminarily investigated by means of the disc \& star light synthesis model 
introduced by \citet{siwak18b} even despite the fact that the basic assumptions of this methodology 
-- simultaneous space-based coverage to determine accurate values of distinct families of QPOs 
and ground-based, multi-colour observations to determine spectral properties of these QPOs -- is not fulfilled.
Instead, we rely on possible values of QPOs inferred from our sparse ground-based 
data (Sec~\ref{g-bvar}) and we also assume that they occurred over the full two-to-three-week-long monitoring runs.

\subsubsection{Update of disc parameters}
\label{update}

\begin{figure*}
\includegraphics[width = 1\linewidth]{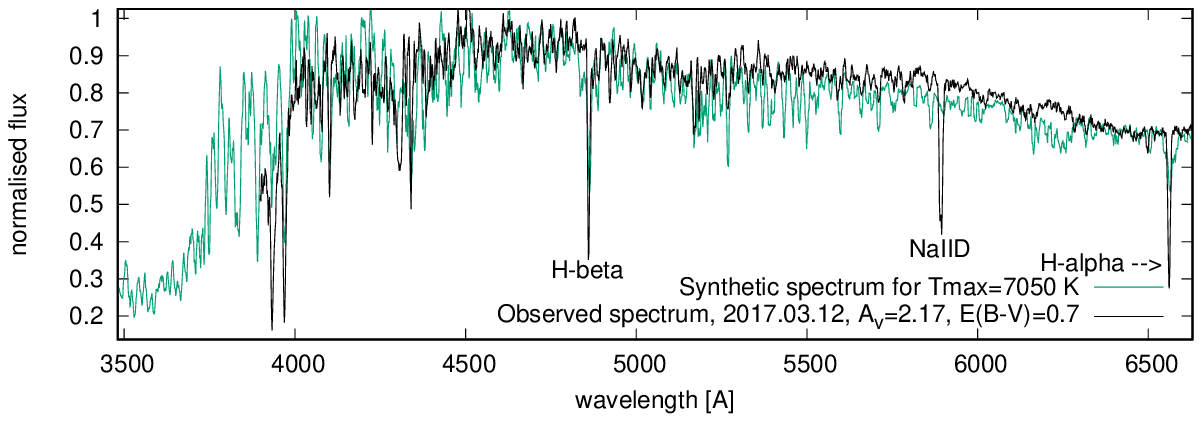} 
\includegraphics[width = 1\linewidth]{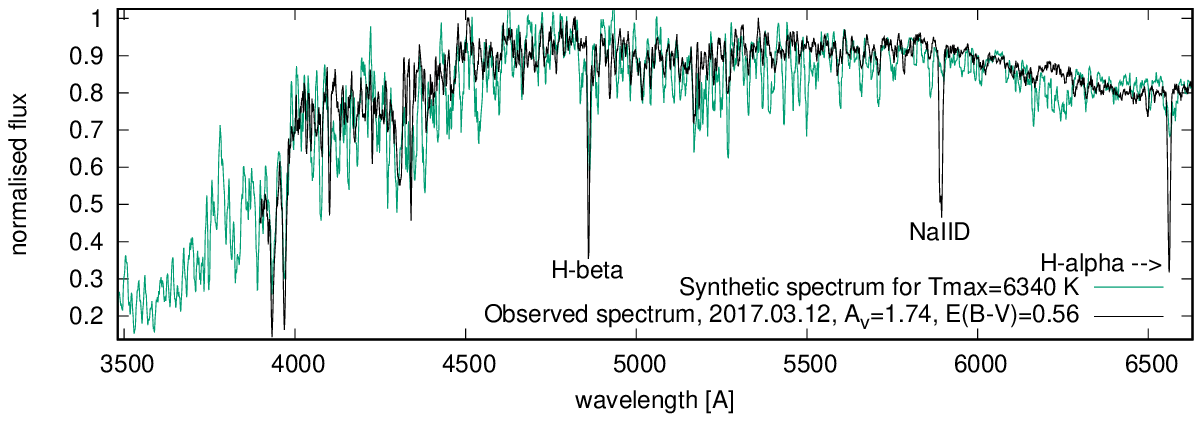}
\caption{Comparison of spectrum observed on March 12, 2017, with two synthetic spectra. 
The observed spectrum was dereddened using an appropriate set of $A_V$ and $E(B-V)$ values corresponding 
to the assumed maximum disc temperature $T_{max}$.}
\label{Fig.spec_comp}
\end{figure*}

In these calculations, we start from the approximate disc parameters of V646~Pup derived 
by \citet{zhu08} (consistent with those derived by \citealt{green06}), including inclination (50~deg), 
$M {\dot M} = 8.1\times10^{-5}$~M$_{\sun}^{2}$~yr$^{-1},$ and inner disc radius $R_{inn}=4.6$~R$_{\sun}$, 
which could be close to the equatorial radius of the bloated star as well, and where 
$\dot M$ is the mass accretion rate transferred from the disc onto the star.
Using these values, we derive maximum disc temperature $T_{max}=7050$~K. 
However, owing to the disc brightness decrease since 1999 based on $\Delta V=0.45$~mag (Fig.~\ref{Fig.longterm}a), 
the flux in visual bands currently amounts to 0.66 of the previous value. 
Assuming that this is only due to mass transfer rate decrease, we currently estimate   
$M {\dot M} = 5.3\times10^{-5}$~M$_{\sun}^{2}$~yr$^{-1}$ and $T_{max}=6340$~K. 
Following the procedure described in Appendix~B of \citet{siwak18b}, 
we obtain $E(B-V)=0.56$~mag and $A_V=1.74$~mag for $R_V=3.1$, which is consistent with $A_V=1.6\pm0.2$~mag 
obtained by \citet{connelly18}.\\
Assuming $T_{max}=6340$~K, we obtained a slightly better fit to standardised 
(in 2013/2014) colour indices: the averaged observed values of $(V-R_c)\approx0.75$ 
and $(V-I_c)\approx1.50$~mag are reproduced by the model at 0.75 and 1.64~mag, respectively. 
The model with $T_{max}=7050$~K results in 0.75 and 1.69~mag, respectively. 
Although the model was not intended to reproduce the disc spectrum in all its specific details, 
in Fig.~\ref{Fig.spec_comp}, 
we compare the observed (corrected for respective values of reddening) and the synthetic disc spectra 
calculated for both values of $T_{max}$. 
All these spectra were normalised to unity at 4500~\AA. 
The synthetic spectra were also smoothed with a 8~\AA ~boxcar to approximately match 
the spectrograph resolution. 
It is evident that the model spectrum with lower disc temperature appears to be a better fit to 
the observed spectrum, especially in the ultraviolet and the blue parts.

We note that our synthetic spectrum shows metallic lines that are deeper, and Balmer lines that 
are more shallow, than observed. 
The first discrepancy is mostly due to unaccounted rotational broadening and the choice of spectral intensities 
for ordinary supergiant stars, while the second is due to increased absorption by a massive wind, 
and perhaps even an  hot boundary layer that has been unaccounted for. 
We do not investigate these problems in full as they have secondary effects on light synthesis results 
in broad-band filters, especially as the current analysis is intended as a guide for further detailed studies. 
The more serious allegations have to be made with regard to the (inevitable at this stage) staggering assumptions, 
starting from the uncertainty during the flux-calibration procedure, through a rough estimation 
of the current $M {\dot M}$ value (apart from the fact that it was earlier estimated using $A_V=2.2$~mag) 
to the simple choice of the $R_V$ parameter, which leads to a maximum disc temperature uncertainty 
as large as 500~K. 
Thus, the results obtained in this work must be treated with a caution until new precise measurements are obtained 
in the broad spectral range at the same time in order to properly update spectral energy distribution 
\citep{abraham04, green16} and relevant disc parameters can be established. 
This may impact the value of $R_{inn}$ as well, as the new Gaia distance of 1093~pc \citep{gaia18} suggests a disc 
luminosity of 102~L$_{\odot}$. 

% ----------------------- Fig.8: comparison of observed and synthetic spectra ------------------------
%----------------------------------------------------------------------------------------------------- 

\subsubsection{Modelling results}

To reproduce the best-defined slopes observed in colour-magnitude diagrams, we used the model 
with $T_{max}=6340$~K only. 
Following the procedure described in Sec.~3.5.2 in \citet{siwak18b}, we parameterised the disc inhomogeneity 
using local deviations $\Delta T$ from the effective temperature 
$T_{eff}$ being a function of disc radius $R$ in the case of stationary accretion \citep{pringle81}:
\begin{equation}
\label{teff-r} 
T_{eff}^{4}(R)= \frac{3GM{\dot M}}{8{\pi}{\sigma}R^{3}} \left[1-\left(\frac{R_{inn}}{R}\right)^{\frac{1}{2}}\right],
\end{equation}
where $G$ is the universal gravitation constant and $\sigma$ is the Boltzmann constant, while 
$\Delta T \equiv |T(R)-T_{eff}(R)|/T_{eff}(R)$. 
This can be written as follows:
\begin{equation}
\label{inhom}
T(R)= \left\{ \begin{array}{ccc} 
T_{eff}(R) & \mbox{for} & R_{inn} \leqslant R < R_{pert}^{inn}, \\  
                      & & 0 \leqslant \varphi < 2\pi \\                  
(1 + \Delta T)\times T_{eff}(R) & \mbox{for} & R_{pert}^{inn} \leqslant R \leqslant R_{pert}^{out},\\
                     & & 0 \leqslant \varphi < \pi \\
                     T_{eff}(R) & \mbox{for} & R_{pert}^{inn} \leqslant R \leqslant R_{pert}^{out},\\ 
                     & & \pi \leqslant \varphi < 2\pi\\
 T_{eff}(R) & \mbox{for} & R > R_{pert}^{out},\\
                     & & 0 \leqslant \varphi < 2\pi,\\
\end{array}\right. 
\end{equation}
where $R_{pert}^{inn}$ and $R_{pert}^{out}$ define the inner and outer radius of a disc ring, 
in which the local effective temperatures $T(R)$ deviate by $\Delta T$ from these predicted 
by Equation~\ref{teff-r}, 
while $\pi$ is an azimuthal angle $\varphi$ chosen a priori, which determines the azimuthal width 
of the disc inhomogeneity. 
% I DECIDED TO SKIP THIS LONG SENTENCE as it brings only little to the context:
%We already tested this calculation for FU~Ori by estimating the azimuthal and radial size 
%of the inhomogeneous disc area 
%(contained between $R_{pert}^{inn}$ and $R_{pert}^{out}$) for a set of small or moderate $\Delta T$, 
%and turning the disc rotation on; it is then possible to obtain synthetic amplitudes and colour index 
%variations that are consistent with observations. 

We searched the parameter space manually with a step of 0.01 in $\Delta T$.
The same $\Delta T$ is always assumed for all filters. 
A step of $\Delta R=1$~R$_{\sun}$ is used to estimate $R_{pert}^{inn}$, $R_{pert}^{out}$ as well 
as an optimal radial width of the hot spot on the disc $R_{pert}^{out}-R_{pert}^{inn}$.\\
Using this procedure, we were unable to reproduce the strong correlations present in 2013 
colour-magnitude diagrams assuming disc inhomogeneities parameterised by moderate and large $\Delta T$.
We suspect that these light changes could be quasi-periodic ($\sim1-2$~d) for at least a fraction of this run, 
which remains invisible for us due to the 1~d sampling and limited photometric accuracy. 
Such strong correlations could then indeed be produced if these variations would occur at the inner boundary 
of the disc or, perhaps, on the surface of the star, and could be explained by a non-uniform shape 
of the hot transition region, as considered for a similar one week-long event observed in FU~Ori 
during {\it Segment~II} in \citet{siwak18b}. 
Unfortunately, V646~Pup was too faint for the 50-cm telescope to be observed in blue and ultraviolet band.\\
Contrary to the above result, the 2016 colour-index variations (i.e. the year we observed signs 
of $V-(B-V)$ colour reversal) could be explained by the rotation of a hot plasma bubble localised between 
15-17~R$_{\odot}$, parameterised by $\Delta T=0.16$. 
Assuming the 7-8~d quasi-period (Sec.~\ref{g-bvar}) and Keplerian rotation of the disc, 
we can estimate the stellar mass at 0.7-0.9~M$_{\odot}$. 
We note that clear evidence for colour index reversals was often observed in quiet CTTS's discs 
in infrared light (\citealt{gunther14, wolk18} and references in these papers) and explained 
by significant changes in the inner disc structure.\\
To reproduce the 2017 colour-magnitude diagrams, we first calibrated our model to Sloan system 
using Vega magnitudes in $AB_{95}$ system in Table~7 of \citet{fukugita96}. 
Linear limb darkening coefficients for Sloan system were adopted from Table~3 of \citet{claret04}.  
We found that these light variations could be owing to the rotation of disc inhomogeneity parameterised by 
$\Delta T= 0.15$ and localised at 12-15~R$_{\odot}$. 
The Keplerian radius of the associated 5-6~d quasi-period (Sec.~\ref{g-bvar}) is equal to 13.5~R$_{\odot}$ 
for a 0.9~M$_{\sun}$ star.

This preliminary investigation suggests that longer periods arise at more distant parts 
of the disc, similarly as in FU~Ori \citep{siwak18b}. 
Multiple coordinated observing runs are still necessary to validate this preliminary result.

% ----------------------- Fig.9: Broadening Functions - short timescales ------------------------------
\begin{figure*}
\includegraphics[width=0.5\linewidth]{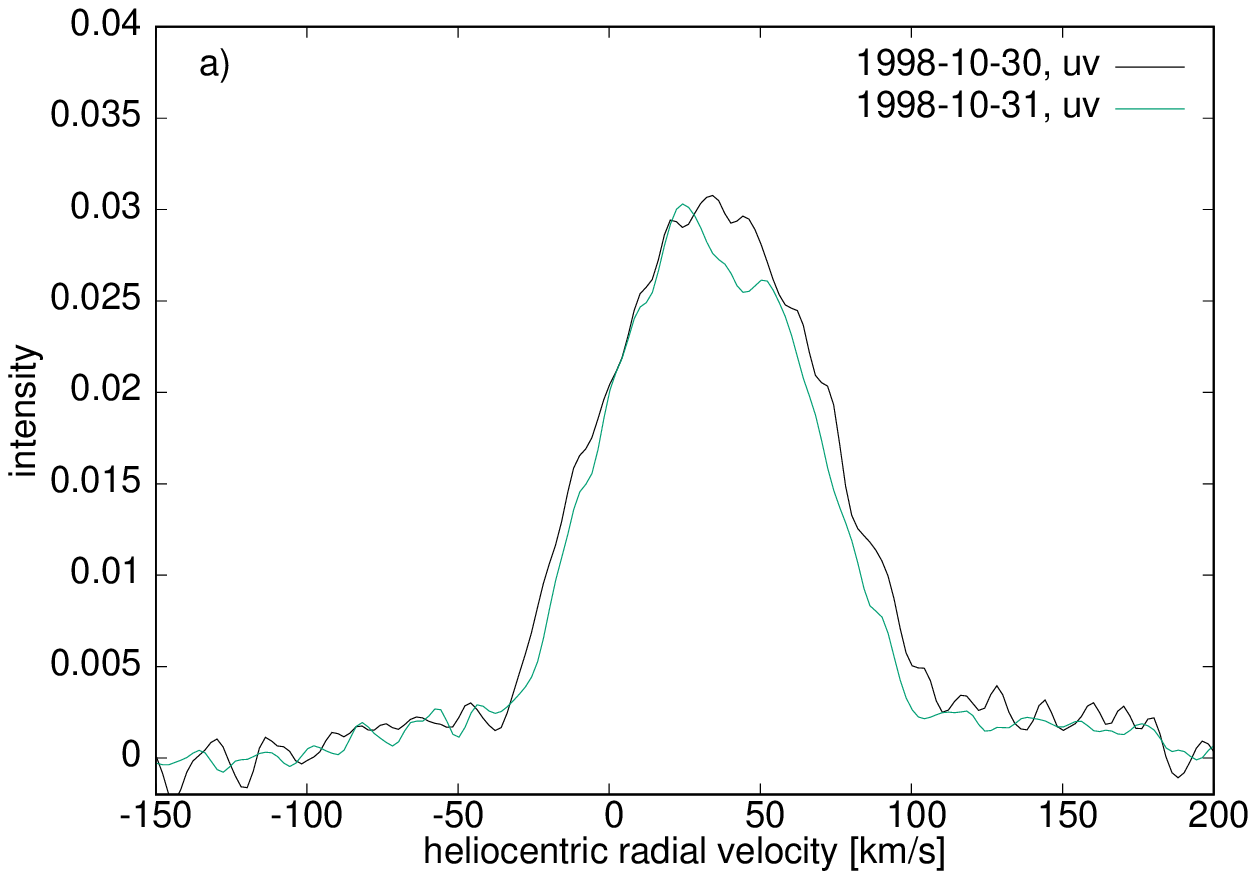}
\includegraphics[width=0.5\linewidth]{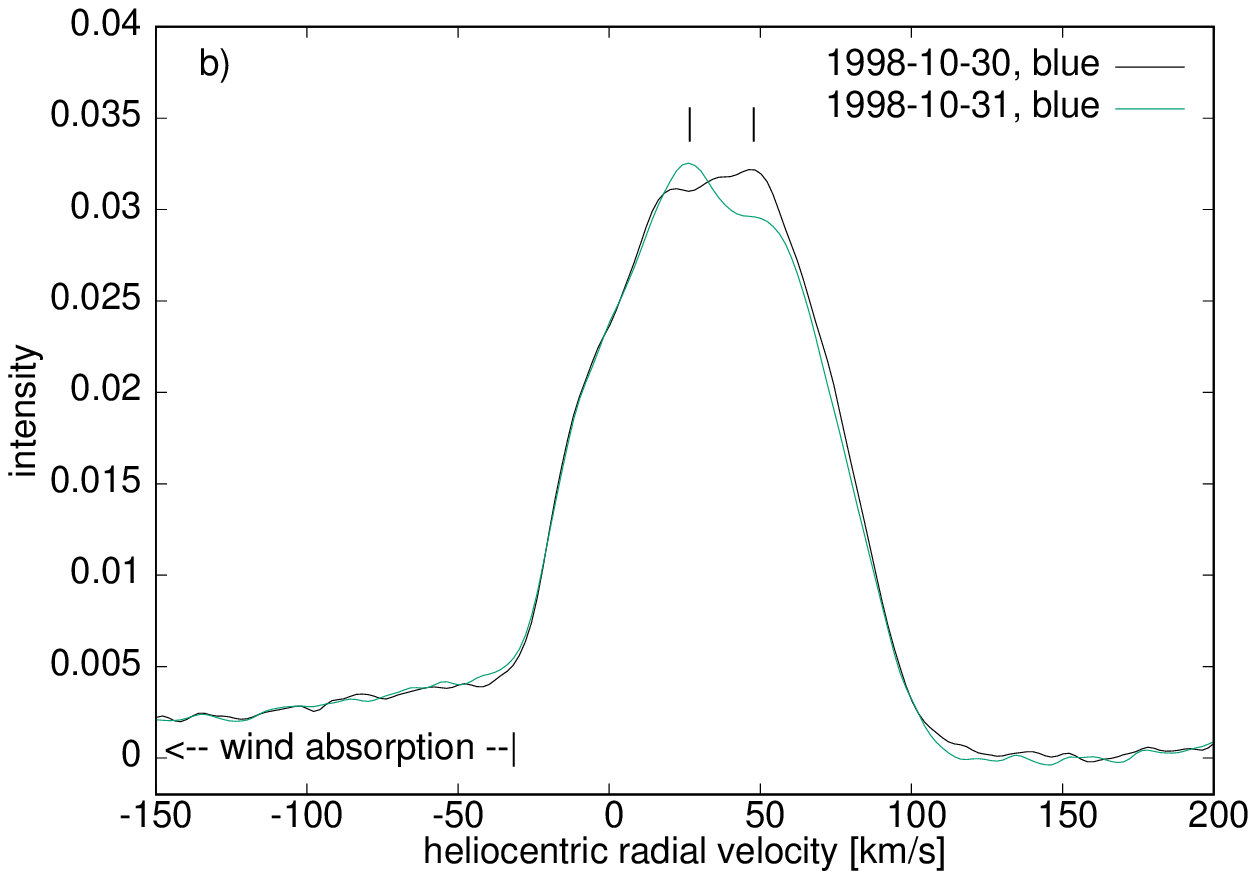}\\
\includegraphics[width=0.5\linewidth]{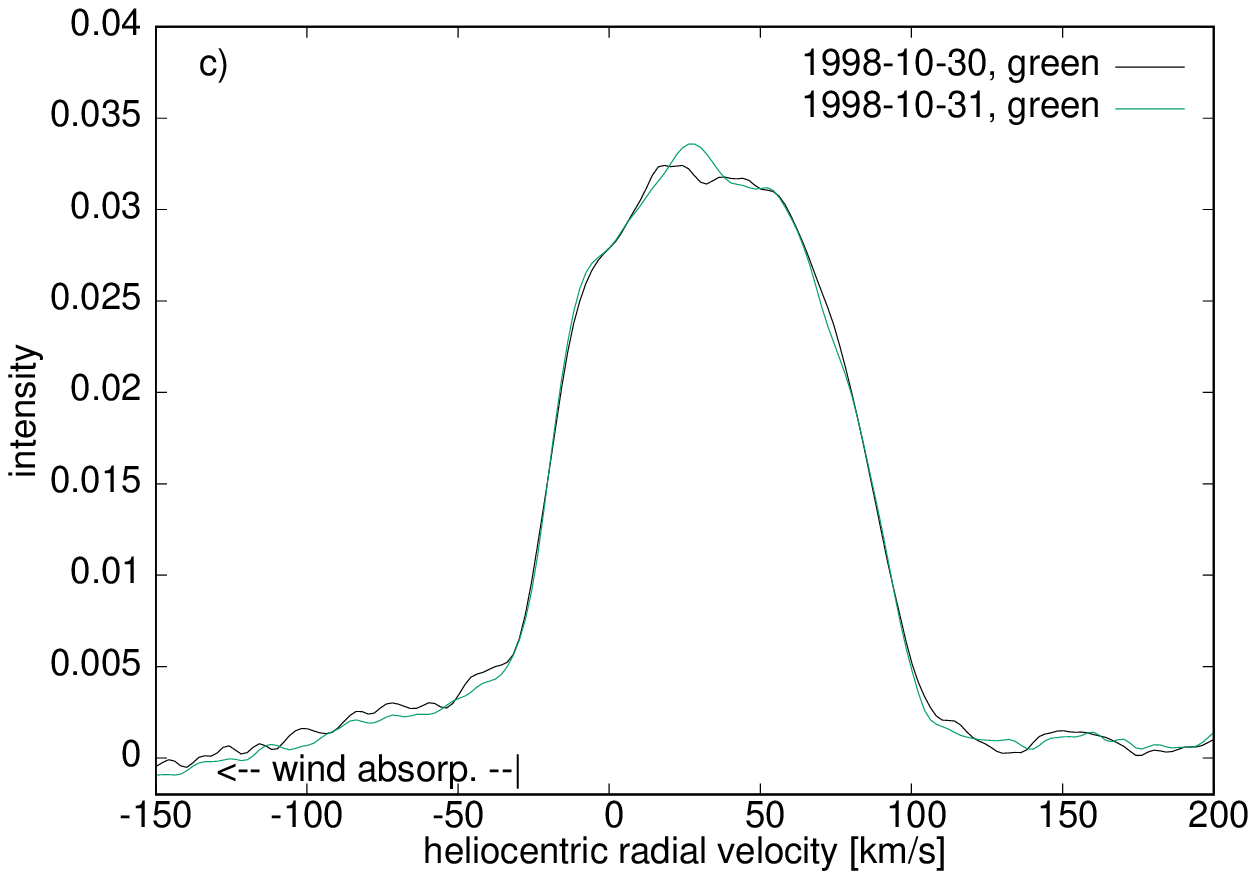}
\includegraphics[width=0.5\linewidth]{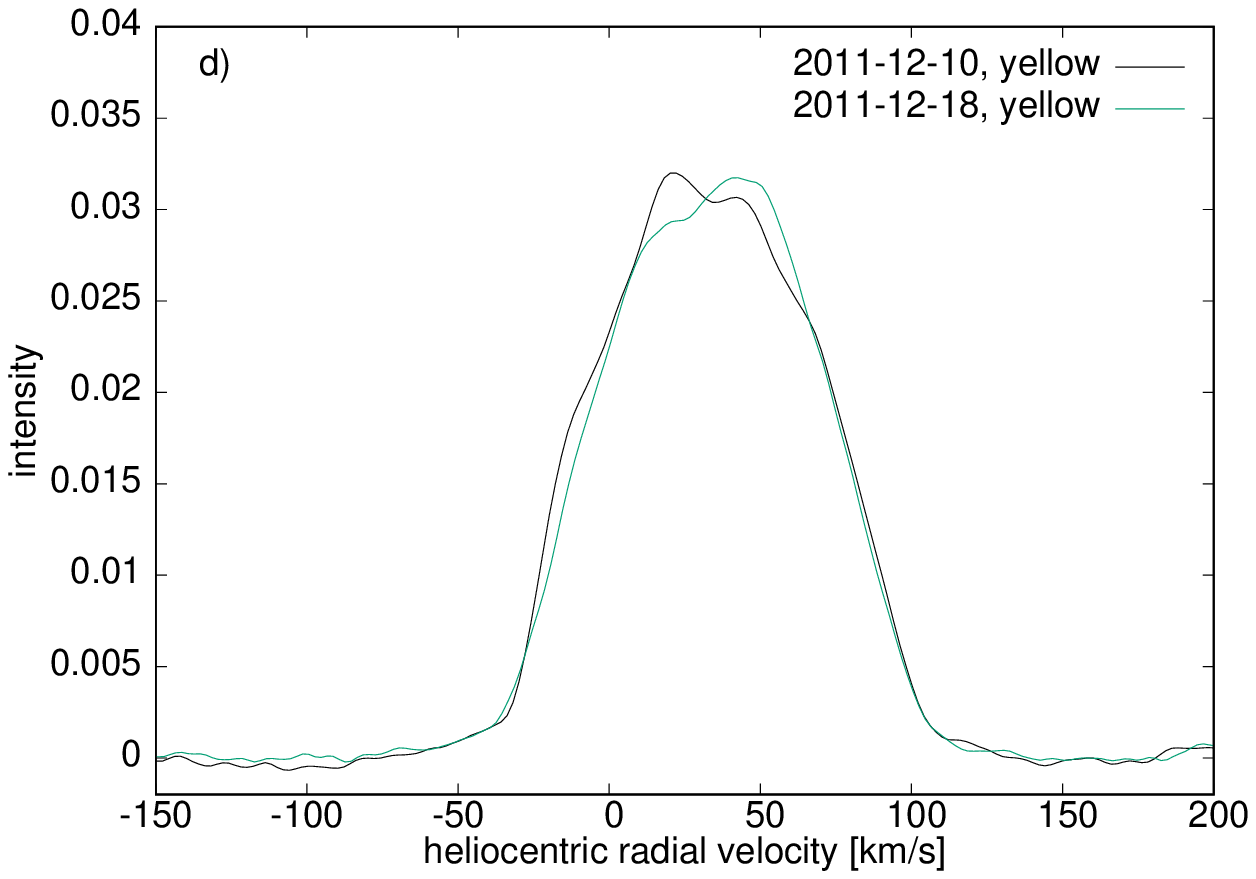}
\caption{Changes of the disc rotational profiles occurring on shorter time-scales.} 
\label{Fig.hires1}
\end{figure*}
%----------------------------------------------------------------------------------------

% ----------------------- Fig.10: Broadening Functions - longer timescales ------------------------------
\begin{figure*}
\includegraphics[width=0.33\linewidth]{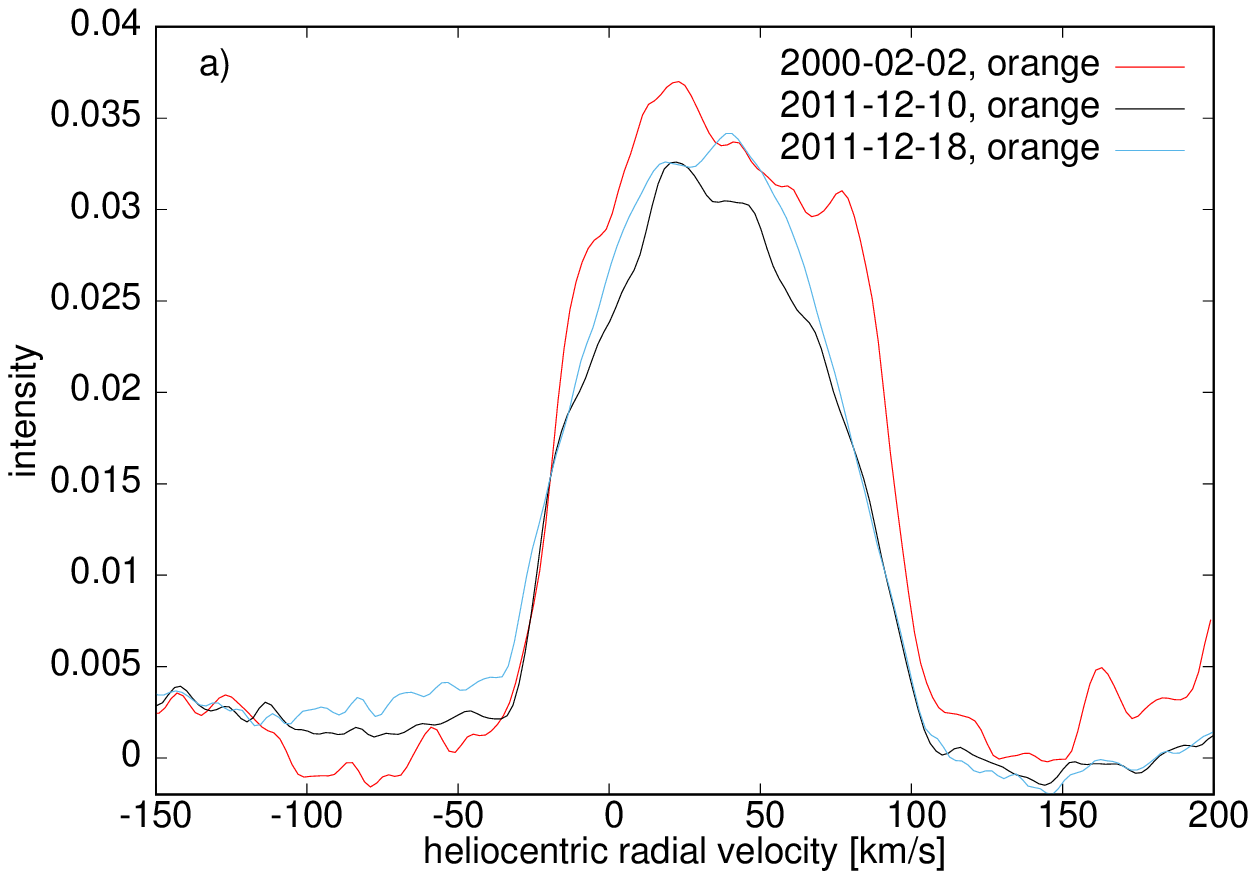}
\includegraphics[width=0.33\linewidth]{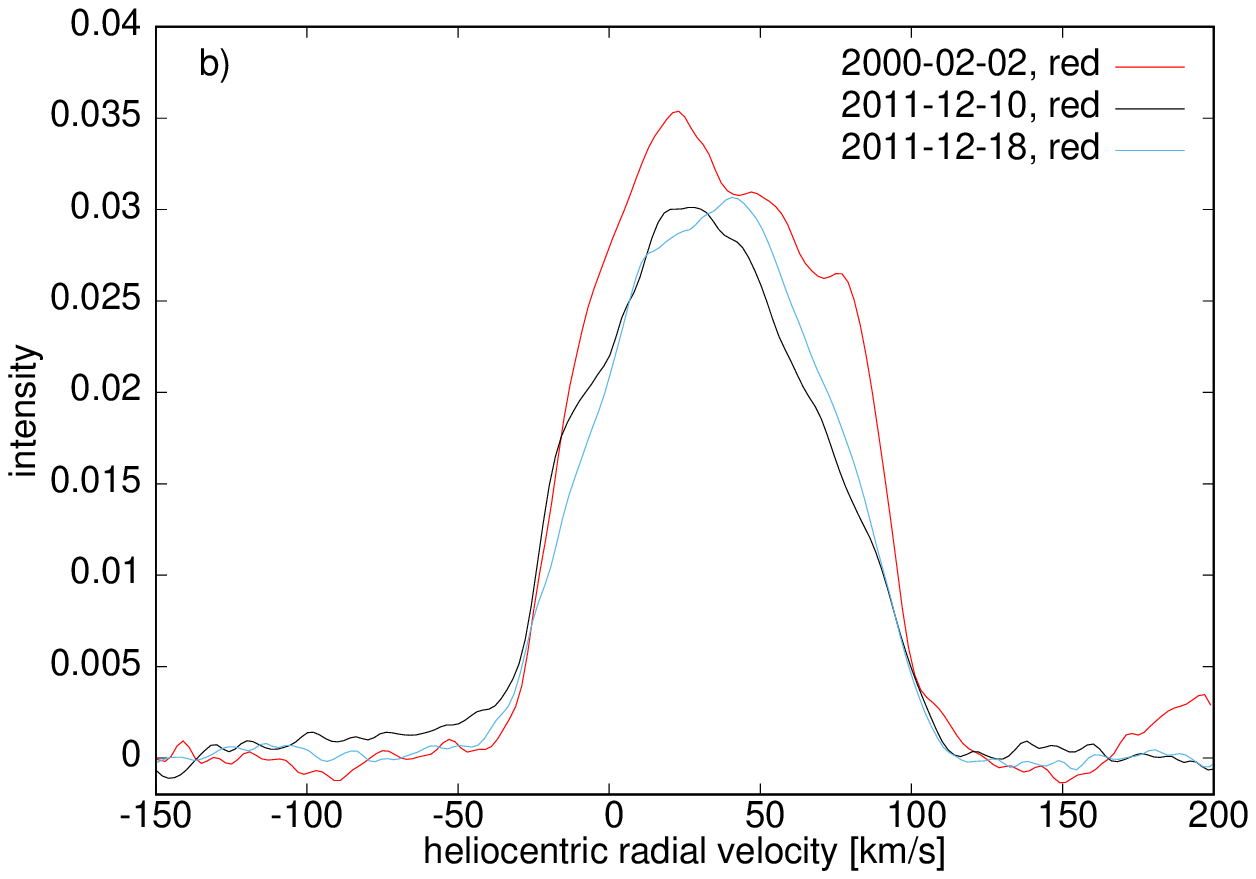}
\includegraphics[width=0.33\linewidth]{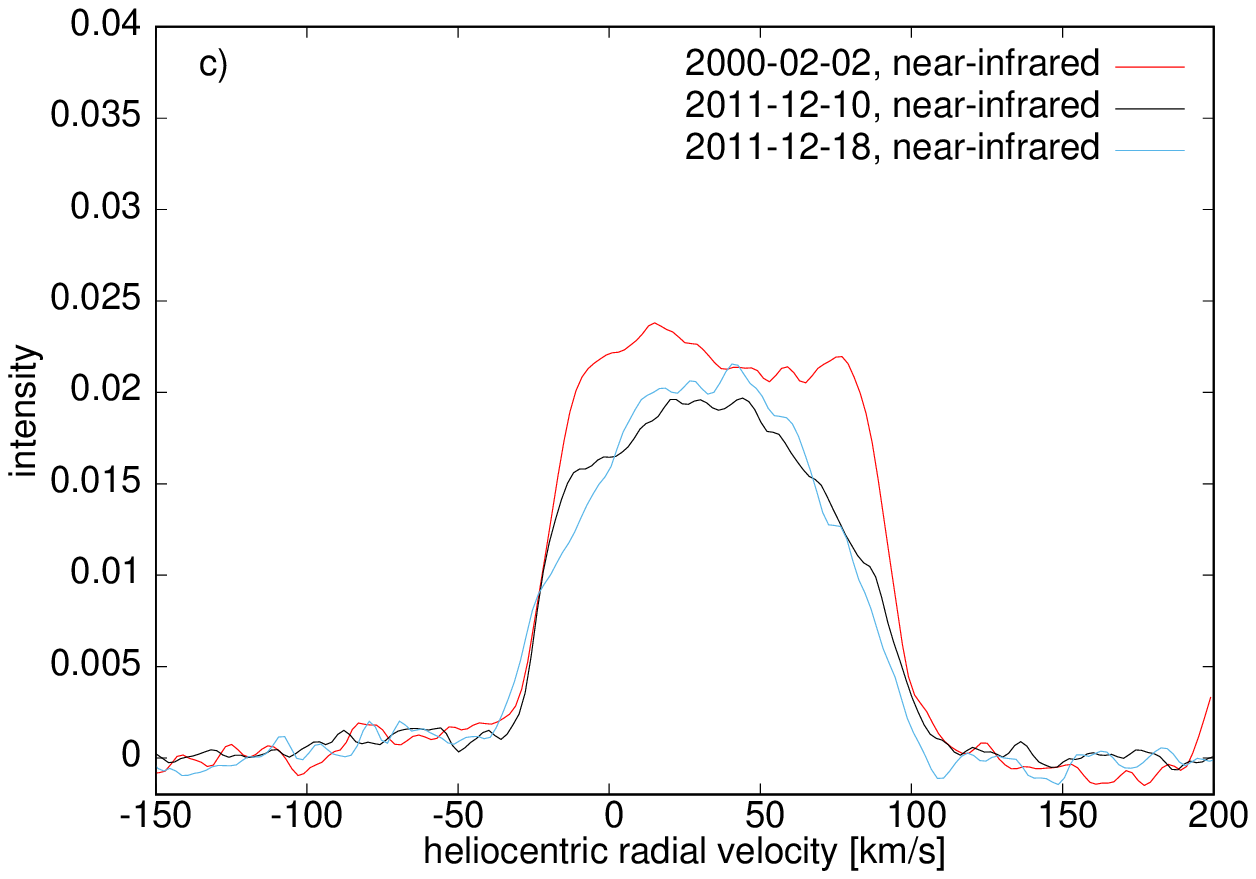}
\caption{Changes of the disc rotational profiles occurring on longer time-scales.} 
\label{Fig.hires2}
\end{figure*}
%----------------------------------------------------------------------------------------

\subsection{Rotational profiles of the disc}
\label{spec-keck}

The broadening function (BF) method was originally introduced for studies of close binaries in the early 1990s 
and then significantly improved (\citealt{ruc12}, and references therein). 
It determines the Doppler broadening kernel in the convolution 
equation transforming a sharp-line template into the observed spectrum. 
As a result, any radial velocity space induced effects are isolated as the kernel of the transformation. 
The time series of broadening functions can also be used for determination of rotational periods of stars 
with photospheric inhomogeneities, such as spotted stars as with CTTS \citep{siwak16}, and for accretion disc mapping. 
Given the obvious advantages of the BF over the cross-correlation function (CCF) method previously applied 
for weak metallic lines studies in other FUors \citep{hartmann85, hartmann87, welty92, herbig03, powell12}, especially with regard to
the linearity and increased resolution, we decided to apply it for the first time for a FUor.

In order to calcuate the BFs for V646~Pup, we transformed all spectra to radial velocities binned with 
the step of $\Delta v=2.0$~km~s$^{-1}$, which is the same for all echelle orders. 
We extracted the BFs from spectral regions free of broad Balmer and metallic lines affected by strong 
disc wind features (e.g. a magnezium triplet, see e.g. \citealt{herbig03}). 
Synthetic spectrum of a supergiant star calculated with the step of 0.01~\AA~ (6250~K, $\log g=2.5$, 
\citealt{gray10}) was used as the template during the deconvolution. 
For BF extraction beyond 6800~\AA, we used a similar spectrum from the POLLUX database \citep{palacios10}. 
The BFs were later smoothed with a gaussian of $1.5\times \Delta v$ width to match the effective spectrograph 
resolution of 7~km~s$^{-1}$.
Furthermore, we averaged BFs obtained in 4-12 consecutive echelle orders to get 'mean BFs' 
in five bands, defined as: uv (3611-4077), blue (4067-4559) and green (4596-4845) 
for 1998 data, and orange (5667-6094), red (6058-6650) and near-infrared (6629-8125~\AA) 
bands for the 2000 and 2012 data. 
In addition, the 2012 data alone allow us to define the yellow band (4776-5992~\AA). 
We present our obtained results in Fig.~\ref{Fig.hires1} and Fig.~\ref{Fig.hires2}. 
Some bands are unavailable for certain years or they are combined using different number of echelle orders 
due to different spectrograph setups. 
The BFs were corrected for relative velocity of the observer with respect to the Sun, but they were 
intentionally left uncorrected for the mean velocity ($33.1\pm2.5$~km~s$^{-1}$ obtained from gaussian 
fits to BFs) owing to band- and time-dependent asymmetries. 

In accordance with the results obtained in the past for a few other FUors, V646~Pup does not show purely 
double-peaked rotational profiles. 
A structure that most closely resembles such a pattern was only present in 2000 but then became invisible in 2011, 
even at longer wavelengths. 
Such a rounded shape for CCF profiles was explained in the past as having been caused by a wind shell, manifesting 
itself as an excess of absorption (i.e. in CCF or BF intensity) mostly near the blue rotational peak 
(e.g. \citealt{hartmann85, welty92}, and references in these papers). 
\citet{herbig03} proposed that lack of double-peaked profiles could also be due to a contribution from 
a fast-rotating bloated central star. 
%Except for the above cases, wind absorption signatures 
%in the blue peak do appear to be present in V646~Pup, especially in the 2000 data (Fig.~\ref{Fig.hires2}).  
In addition, despite our attempts to avoid metallic lines that have obviously been contaminated by the wind, in 
blue and green continua (1998) we observe a tail extending at least to $-180$~km~s$^{-1}$ with respect 
to the profile center, which is of the same nature as that decomposed in detail by \citet{herbig03}.\\ 
The most important result of this analysis is, however, that the BFs show well-defined (i.e. higher than BF errors, 
determined by scatter in the surrounding continua) excess of intensity on top and on the edges 
(Fig.~\ref{Fig.hires1}abcd). 
This excess is indicated by short marks in Fig.~\ref{Fig.hires1}b only, and is present on positive velocities 
as often as on the negative ones, which suggests that it is more reasonable to attribute it to a rotating 
disc inhomogeneity; furthermore, we note that it changes its position with respect to the star sometimes 
(in 1998) within one day. 
The BFs indicate on wavelength-dependent character of these inhomogeneities: for instance, 
the 1998 data show largest differences in the uv and blue bands and only small ones in the green. 
Future coordinated spectroscopic and photometric observations will make it possible to validate this assumption.

Aside from the above, the orange, red, and infrared BFs show evidence of an intensity decrease over 
the first decade of the 21st century (Fig.~\ref{Fig.hires2}abc). 
To examine whether this is an effect of a weakening outburst (i.e. disc luminosity), we first aligned the continua 
of all BFs to zero level and integrated the intensities over wavelengths. 
Integrals calculated from the orange and the red band BFs in December, 2011, are by 1.25 smaller than those 
obtained in February, 2000. 
Linear properties of the BF method allow us to translate this value to a $0.22\pm0.03$~mag  drop in brightness. 
This is in accordance with the decline rate obtained from the fit to Johnson-V observations 
(Fig.~\ref{Fig.longterm}a), indicating at 0.21~mag, a weakening over the 11.8 years between these observations. 
It is also noteworthy that the most significant intensity drops are evident in the wings of the BFs, 
that is, close to the star itself. 
This finding independently confirms the innermost disc temperature decrease, as shown in Sect.~\ref{update}.  

\section{Summary and conclusions}
\label{summary}

In this paper, we analyse recent uncoordinated space- and ground-based photometric and spectroscopic observations of V646~Pup 
in order to detect and offer a preliminarily characterisation of low-amplitude light variations superimposed 
on the general decreasing trend in brightness. 
This small-scale variability carries important information about the physical processes taking place in the inner 
disc and potentially also on the star itself during the stage of episodic accretion.
 
New $V$-filter data confirm the gradual light decrease in the disc, which was previously noticed by \citet{reipurth02}. 
Assuming that the decline rate 0.0175~mag~yr$^{-1}$  stays the same, the disc will remain 
in the outburst for another century. 
Because of the 0.5~mag brightness decrease in visual band over 35 years, we also made an attempt to update 
several disc-related physical parameters. 
As a result, we found that the smaller disc temperature (6340~K) and smaller extinction 
and reddening ($A_V=1.74$~mag and $E(B-V)=0.56$~mag, respectively) better match the flux-calibrated 
spectrum obtained in 2017.

The power spectrum of V646~Pup computed from the detrended {\it ASAS-SN} data does not reveal any significant periods 
nor quasi-periods. 
Instead, well-defined quasi-periods are present in the space-based {\it TESS} light curve: its first half 
is dominated by double-peaked light variability pattern, which is surprisingly similar 
to that observed in the 2013-2014 {\it MOST} light curve of FU~Ori. 
We assume that the primary maximum could be caused by changing visibility of a well-defined hot plasma 
parcel in the course of its rotation around the star, whereas the secondary maximum could be due to reflection effect. 
Nonetheless, it is premature to conjecture on this possibility with greater certainty 
due to the lack of simultaneous multi-colour and high-resolution spectroscopic coverage. 
Similar reservations can be addressed to the results obtained from the modelling of colour-magnitudes 
diagrams, which, in turn, were received without space-based support that could allow  specific 
values of associated quasi-periods to be determined. 
Nevertheless, these results are sufficient to support the preliminary conclusion 
that the inner disc in V646~Pup is Keplerian and the stellar mass is of 0.7-0.9~M$_{\odot}$. 
If this is correct, then the double-peaked $4.8\pm0.2$~d variability observed by {\it TESS} in 2019 could 
be caused by rotation of a hot plasma parcel that appeared at 10.5-12~R$_{\odot}$. 
Certainly, such a well-defined hot plasma parcel should also manifest its existence in the disc's 
rotational profiles. 
Broadening function profiles calculated from archival Keck data do indicate such possibility.
 
V646~Pup and FU~Ori appear to be ideal among FUors for testing 
the results obtained from the global three-dimensional magnetohydrodynamical investigation 
of the mechanism leading to and sustaining episodic accretion for over a century by \citet{zhu20}. 
It would be also very instructive to compare the results obtained for these two aging FUors with those 
for the younger ones (V2493~Cyg -- \citealt{semkov10}, V960~Mon -- \citealt{maehara14}, 
Gaia~18dvy -- \citealt{szegedi-elek20}), which can be obtained in future coordinated time-series 
photometric and high-resolution spectroscopic observations.

\section*{Acknowledgments}
\label{thanks}
MS and WO are grateful to the Polish National Science Centre for the grant 2012/05/E/ST9/03915. 
This project has received funding from the European Research Council (ERC) under the European 
Union's Horizon 2020 research and innovation programme under grant agreement No 716155 (SACCRED).\\
This paper includes data collected by the {\it TESS} mission. 
Funding for the {\it TESS} mission is provided by the NASA Explorer Program. 
This study was based on observations made at the Cerro Tololo Inter-American 
Observatory with the 0.9-m telescope operated by the {\it SMARTS} Consortium, 
and observations made at the South African Astronomical Observatory. 
Polish participation in SALT is funded by grant No. MNiSW DIR/WK/2016/07.\\
The data presented herein were obtained at the W. M. Keck Observatory, which is 
operated as a scientific partnership among the California Institute of Technology, 
the University of California and the National Aeronautics and Space Administration 
within the programmes ID H22aH and H17aH (George Herbig), C186Hr (Lynne Hillenbrand) and H254H6 (Bo Reipurth). 
The Observatory was made possible by the generous financial support of the W. M. Keck Foundation.
This research has made use of the Keck Observatory Archive (KOA \#79971), which is 
operated by the W. M. Keck Observatory and the NASA Exoplanet Science Institute (NExScI), 
under contract with the National Aeronautics and Space Administration.\\ 
This paper made use of NASA's Astrophysics Data System (ADS) Bibliographic 
Services, operated by the Smithsonian Astrophysical Observatory under NASA Cooperative 
Agreement NNX16AC86A.\\ 
This research was achieved using the POLLUX database ( http://pollux.oreme.org )
operated at LUPM  (Universit\'e Montpellier - CNRS, France) with the support of the PNPS and INSU.\\
This research has made use of the SIMBAD database, operated at CDS, Strasbourg, France.\\
We acknowledge Dr. Francois van Wyk and the entire {\it {\it SAAO}} staff 
for their hospitality, and Dr. Jennifer G. Winters for help in efficient start of the {\it CTIO} run. 
Special thanks are also due to an anonymous referee for highly useful suggestions
and comments on the previous version of the paper.

%\begin{appendix}
%\section{Light curves}
%\label{lc_app}
%The tables Tab.A1 - Tab.A8 with light curves and flux-calibrated spectrum (see in Section~\ref{observations}) 
%are available in electronic form only at the CDS via anonymous 
%ftp to cdsarc.u-strasbg.fr (130.79.128.5).
%\end{appendix}
\end{document}